\documentclass[11pt]{article}
\pdfoutput=1
\usepackage{fullpage}
\usepackage{graphicx}
\def\lsim{\mathrel{\rlap{\lower4pt\hbox{\hskip1pt$\sim$}}
    \raise1pt\hbox{$<$}}}         
\def\gsim{\mathrel{\rlap{\lower4pt\hbox{\hskip1pt$\sim$}}
    \raise1pt\hbox{$>$}}}         
\begin{document}
\begin{flushright}
INT PUB 08-26
\end{flushright}
\begin{center}
{\bf NEUTRINO ASTROPHYSICS} \\
 W. C. Haxton \\
 Institute for Nuclear Theory and Department of Physics \\
 Box 351550 University of Washington, Seattle, WA 98195 \\
 email: Haxton@phys.washington.edu
\end{center}

\section{Introduction}
\noindent
The neutrino \cite{hh00} is an elementary particle that scatters only through the weak interaction, and consequently rarely interacts in matter.  Neutrinos are neutral, carry spin-1/2, and are members of the family of elementary particles called leptons.  Thus they differ from the quarks of the standard model (spin-1/2 particles which participate in strong and electromagnetic interactions) and from the other leptons, which are charged and thus interact electromagnetically.   Neutrinos 
and their antiparticles come in three types $-$ or flavors $-$ labeled according to the charged partners, the electron, muon, or tauon, that accompany neutrino production in charge-changing weak interactions.  The most familiar such reaction is $\beta$ decay
\[ (N,Z) \rightarrow (N-1,Z+1) +  e^- + \bar{\nu}_e \]
in which a nucleus containing N neutrons and Z protons decays to a lighter nucleus by converting
a neutron to a proton, with the emission of an electron and an electron antineutrino.  Indeed, it was the apparent absence of energy conservation in nuclear $\beta$ decay that first lead Wolfgang Pauli, nearly 80 years ago, to suggest that some undetected neutral particle (the $\nu_e$) must be escaping from nuclear $\beta$ decay experiments.

Neutrinos play a very special role in astrophysics \cite{bahcallbook}.  First, they are the direct byproducts of the nuclear reaction chains by which stars generate energy: each solar conversion of four protons into helium produces two neutrinos, for a total of $\sim$ 2 $\times$ 10$^{38}$ neutrinos each second.  The resulting flux is observable on Earth.  These neutrinos carry information about conditions deep in the solar core, as they typically leave the Sun without further interacting.   They also provide experimentalists with opportunities for testing the properties of neutrinos over long distances.  Second, they are produced in nature's most violent explosions, including the Big Bang, core-collapse supernovae, and the accretion disks encircling supermassive black holes.  Recent discoveries of neutrino mass show that neutrinos comprise a small portion of the ``dark matter" that influences how large-scale structure -- the pattern of voids and galaxies mapped by astronomers -- formed over cosmological times.  Third, they dominate the cooling of many astrophysical objects, including young neutron stars and the degenerate helium cores of red giants.  Neutrinos can be radiated from deep within such bodies, in contrast to photons, which are trapped within stars, diffusing outward only slowly.  Finally, neutrinos are produced in our atmosphere and elsewhere as secondary byproducts of cosmic-ray collisions.  Diagnosis of these neutrinos can help constrain properties of the primary cosmic ray spectrum.

Neutrinos also mediate important astrophysical processes.  The supernova ``neutrino wind," which blows off the surface of the proto-neutron star just seconds after core collapse, is thought to produce conditions favorable for the synthesis of about half of the neutron-rich nuclear species heavier than iron,
through rapid neutron capture (the r-process).  Neutrinos also directly synthesize certain rare nuclei, such as ${}^{19}$F.

Nuclear and particle physicists are exploiting astrophysical neutrino fluxes to do important tests of the standard model of particle physics.  These tests include neutrino oscillations (the process by which a massive neutrino can be produced in one flavor state but detected later as a neutrino with a different flavor), neutrino decay, neutrino mass effects, and searches for nonzero neutrino electromagnetic moments.

\section{Solar Neutrinos}
\noindent
The first successful effort to detect neutrinos from the Sun began four decades ago.  Ray Davis, Jr. and his collaborators constructed a 650-ton detector in the Homestake Gold Mine, one mile beneath
Lead, South Dakota \cite{davis}.  This radiochemical detector, based on the chlorine-bearing cleaning fluid C$_2$Cl$_4$, was designed to capture about one of the approximately 10$^{18}$  high-energy neutrinos that penetrated it each day -- the rest
passed through the detector, without interacting.  The neutrino-capture reaction was 
inverse-electron-capture
\[ {}^{37}\mathrm{Cl} + \nu_e^{solar} \rightarrow {}^{37}\mathrm{Ar} + e^-. \]
The product of this reaction, ${}^{37}$Ar, is a noble gas isotope with a half life of about one month.  It can be efficiently removed from a large volume of organic fluid by a helium gas purge, then counted in miniature gas proportional counters as  ${}^{37}$Ar decays back to ${}^{37}$Cl.  Davis typically exposed his detector for
about two months, building up to nearly the saturation level of a few dozen argon
atoms, then purged the detector to determine the number of solar neutrinos captured during this period.

Within a few years it became apparent that the number of neutrinos detected was only about one-third that
predicted by the standard solar model (SSM) \cite{BSB,TC}, that is, the model of the Sun based on the standard theory of main sequence stellar evolution. Some initially attributed this ``solar neutrino problem" to uncertainties in the SSM: As the flux of neutrinos most important to the Davis detector vary as $\sim$ $T^{22}_c$, where $T_c$ is the solar core temperature, a 5\% theory uncertainty in $T_c$ could explain the discrepancy.  In fact, the correct explanation for the discrepancy proved much more profound.  Davis was awarded the 2002 Nobel Prize in Physics for the Cl experiment.

\begin{figure}
\begin{center}
\includegraphics[width=12cm]{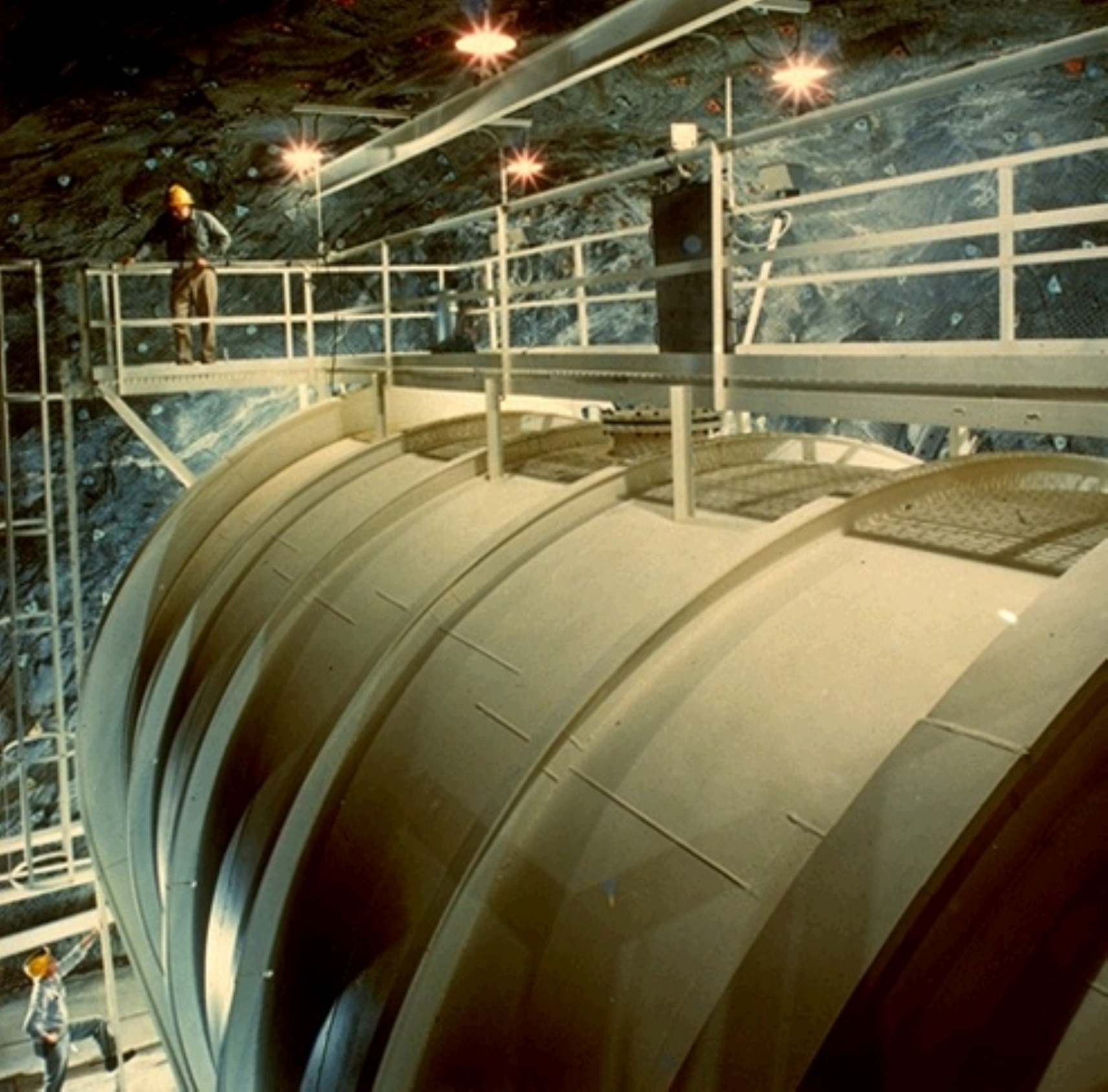}
\end{center}
\caption{ The Homestake Mine's chlorine detector, which Ray Davis Jr. and colleagues
operated for over three decades.}
\label{fig:one}
\end{figure}

During the three-decade period of  ${}^{37}$Cl detector operations, five other solar neutrino experiments were constructed.  The SAGE and GALLEX/GNO  experiments, radiochemical detectors similar to Cl,  but using ${}^{71}$Ga as a target, were designed to measure the flux of neutrinos from
the dominant low-energy branch of solar neutrinos, the pp neutrinos.  The first detector to measure neutrinos event by event, recording neutrino interactions in real time, was the converted proton decay detector Kamiokande.   The detector contained three kilotons of very pure water, with solar neutrinos scattering off the electrons within the water.  Phototubes surrounding the tank recorded the ring of Cerenkov radiation produced by the recoiling relativistic electrons.  Kamiokande measured the high-energy neutrinos most important to the Davis detector, and thus confirmed the deficit that Davis had originally observed.  Finally, a new generation of massive water (Super-Kamiokande) and heavy-water (Sudbury Neutrino Observatory (SNO)) detectors were constructed.  The later, in a most direct way, showed that solar neutrinos were not missing, but rather hidden by a change of flavor occurring during their transit from the Sun to the Earth -- as will be described later in this chapter.

\subsection{The Standard Solar Model}
\begin{figure}
\begin{center}
\includegraphics[width=10cm]{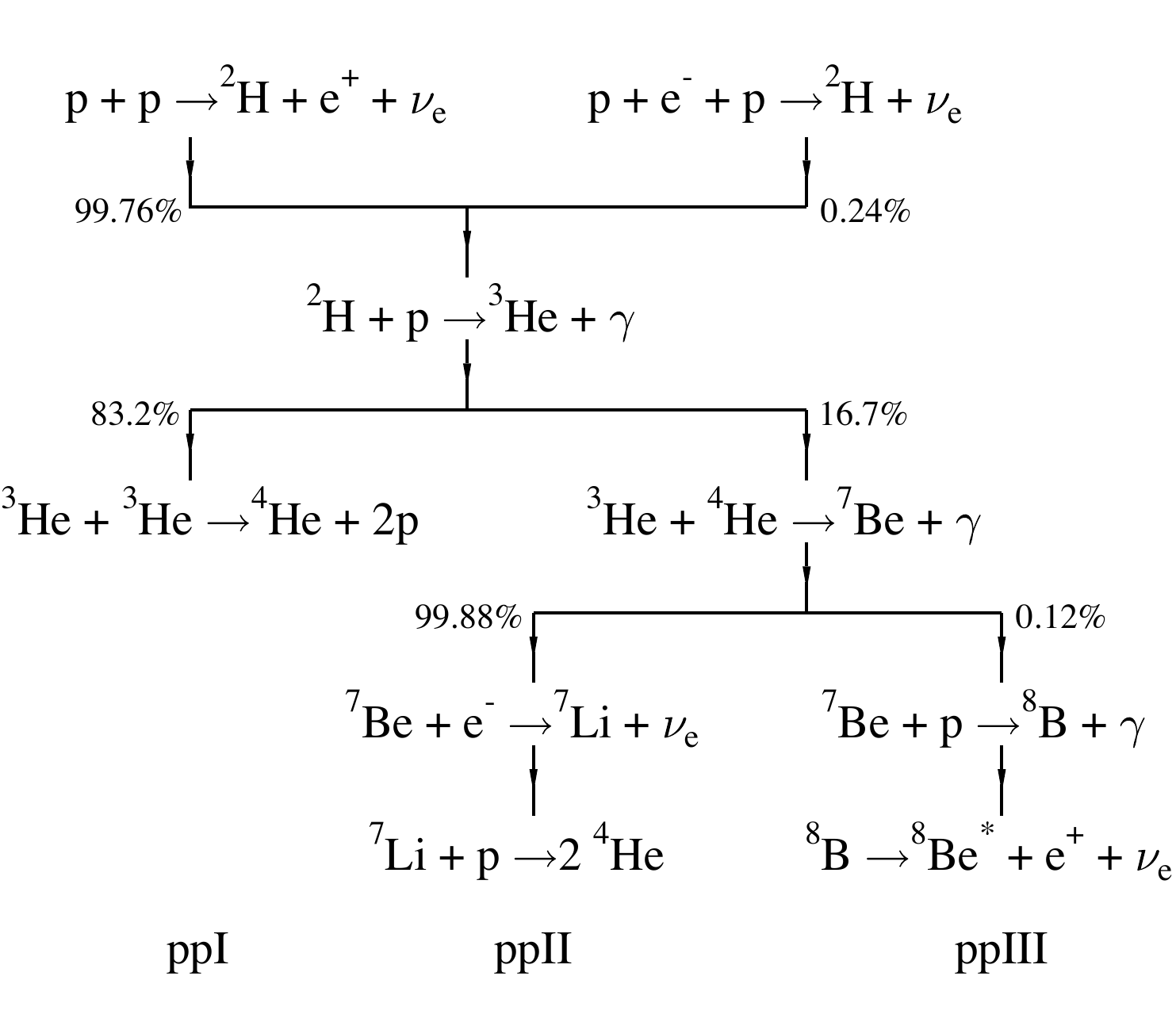}
\end{center}
\caption{ The pp-chain for hydrogen burning.  The relative rates
of competing reactions correspond to the Bahcall-Pinsonneault 2004 SSM \cite{BP}.}
\label{fig:pp}
\end{figure}

The Sun belongs to a class of ``main sequence'' stars that derive their energy from burning protons to He in their cores.  The SSM employs the standard theory of main-sequence stellar evolution, calibrated by the many detailed measurements only possible for the Sun, to follow the Sun from the onset of thermonuclear reactions 4.6 b.y. ago to today, thereby determining the present-day temperature and composition profiles of the solar core.  These profiles govern solar neutrino production and other properties of the modern Sun.  The SSM is based on four basic assumptions:
    
\begin{itemize}
\item The Sun evolves in hydrostatic equilibrium, maintaining a local balance between the gravitational force and the pressure gradient.  To describe this condition in detail, one must  specify the electron-gas equation of state as a function of temperature, density, and composition.
\item Energy is transported by radiation and convection.  While the solar envelope is convective, radiative transport dominates in the core region where thermonuclear reactions take place.  The radiative opacity depends sensitively on the solar composition, particularly the abundances of heavier elements.
\item Thermonuclear reaction chains generate solar energy.
The standard model predicts that 99\% of this energy is produced from the pp chain conversion of four protons into ${}^4$He
(see Fig.~\ref{fig:pp})
\[ 4\mathrm{p} \rightarrow {}^4\mathrm{He} + 2e^+  + 2 \nu_e. \]
The Sun is
a large but slow reactor: the core temperature, $T_c \sim  1.5 \times 10^7$ K, results in typical center-of-mass energies for reacting particles of $\sim$ 10 keV, much less than the Coulomb barriers inhibiting
charged particle nuclear reactions.  Thus reaction cross sections are small: in most cases, as laboratory measurements are only possible at higher energies, cross section data must be extrapolated to the solar energies of interest.
\item The model is constrained to produce today's solar radius, mass, and luminosity.  An important assumption of the SSM is that the Sun was highly convective, and therefore uniform in composition, when it first entered the main sequence.  It is furthermore assumed
that the surface abundances of metals (nuclei with A $>$ 5) were undisturbed by the subsequent evolution, and thus provide a record of the initial solar metallicity.  The remaining parameter is the initial ${}^4$He/H ratio, which is adjusted until the model reproduces the present solar
luminosity after 4.6 billion years of evolution.  The resulting ${}^4$He/H mass fraction ratio is typically 0.27 $\pm$ 0.01, which can be compared to the Big-Bang value of 0.23 $\pm$ 0.01.  
Note that the Sun was formed from previously processed material.
\end{itemize}
  
Three cycles with quite different temperature dependences, reflecting the relative ease or difficulty of Coulomb barrier penetration, comprise the pp chain of Fig.~\ref{fig:pp}.  The competition between the cycles is very sensitive to the solar core temperature $T_c$.  The initial interest in solar neutrinos came from the observation that each of the three cycles is associated with a characteristic neutrino.  Thus, by measuring solar neutrinos, specifically the pp, ${}^7$Be, and ${}^8$B neutrinos, one can determine the relative importance of the ppI, ppII, and ppIII cycles, and consequently determine $T_c$ to an accuracy approaching 1\%.  

The neutrino-producing reactions of the pp chain and CN cycle (a second process for burning protons to He, dominant in massive main-sequence stars, but responsible for only $\sim$ 1\% of solar energy) are summarized in Table \ref{table:one}.  The first six reactions are $\beta$ decays that produce continuous neutrino spectra.   The last two reactions produce line sources of electron capture neutrinos, with widths $\sim$ 2 keV characteristic of the temperature of the solar core.
Table 1 gives the predicted fluxes of these neutrinos at Earth, as determined in two SSM calculations, Bahcall, Serenelli, and Basu (BS05(AGS,OP) \cite{BSB} and Brun, Turk-Chieze, and Morel
(BTCM98) \cite{TC}.  An update of the BS05(AGS,OP) SSM is in progress.

\begin{table}
\caption{Solar neutrino sources and the flux predictions of the
Bahcall, Serenelli, and Basu BS05(AGS,OP) \cite{BSB} and 
the Brun, Turk-Chieze, and Morel BTCM98 \cite{TC} SSMs.
The former uses abundances derived from recent re-analyses of
photospheric absorption lines and new opacities from the Opacity Project.}
\vspace{0.2cm}
\label{table:one}
\begin{center}
\begin{tabular}{|l|c|c|c|}
\hline
 & & & \\
Source & E$_\nu^{max}$  & BS05(ASG,OP) & BTCM98 \\
 & (MeV) & (cm$^{-2}$ s$^{-1}$) &(cm$^{-2}$ s$^{-1}$) \\
\hline
& & & \\
p + p $\rightarrow ^2$H + e$^+ + \nu$ & 0.42 & 6.06$\times 10^{10}$ & 5.98$\times 10^{10}$\\
${}^{13}$N $\rightarrow {}^{13}$C + e$^+ + \nu$ & 1.20 & 2.01$\times 10^{8}$ & 4.66$\times 10^{8}$ \\
${}^{15}$O $\rightarrow {}^{15}$N + e$^+ + \nu$ & 1.73 & 1.45$\times 10^{8}$ & 3.97$\times 10^{8}$ \\
${}^{17}$F $\rightarrow {}^{17}$O + e$^+ + \nu$ & 1.74 & 3.25$\times 10^{6}$ & \\
${}^8$B $\rightarrow {}^8$Be + e$^+ + \nu$ & $\sim$ 15 & 4.51$\times 10^{6}$ & 4.82$\times 10^{6}$ \\
${}^3$He + p $\rightarrow {}^4$He + e$^+ + \nu$ & 18.77 & 8.25$\times 10^{3}$ & \\
${}^7$Be + e$^- \rightarrow {}^7$Li + $\nu$ & 0.86 (90\%) & 4.34$\times 10^{9}$ & 4.70$\times 10^{9}$ \\
 & 0.38 (10\%) & & \\
p + e$^-$ + p $\rightarrow {}^2$H + $\nu$ & 1.44 & 1.45$\times 10^{8}$ & 1.41$\times 10^{8}$ \\
 & & & \\
\hline
\end{tabular}
\end{center}
\end{table}

\subsection{SNO and Super-Kamiokande}
Solar neutrino detection requires the combination of a large detector
volume (to provide the necessary rate of events), very low backgrounds (so
that neutrino events can be distinguished from backgrounds due to cosmic
rays and natural radioactivity), and a distinctive signal.  The first requirement
favors detectors constructed from inexpensive materials and/or materials having
large cross sections for neutrino capture.  The second generally requires
a deep-underground location for the detector, with sufficient rock overburden to
attenuate the flux of penetrating muons
produced by cosmic ray interactions in the atmosphere.  It also requires very careful attention to detector cleanliness,
including tight limits on dust or other contaminants that might introduce
radioactivity, use of low-background construction materials, control of radon, and
often the use of fiducial volume cuts so that the outer portions of a detector
become a shield against activities produced in the surrounding rock walls.

There are several possible detection modes for solar neutrinos,
interesting because of their different sensitivities to flavor.
The early radiochemical experiments using ${}^{37}$Cl and ${}^{71}$Ga
targets were based  on the charged current weak reaction 
\[ \nu_e + (N,Z) \rightarrow e^- + (N-1,Z+1) \]
where the signal for neutrino absorption is the growth over time of 
very small concentrations of the
daughter nucleus $(N-1,Z+1)$ in the detector.  As the spectrum of solar
neutrinos extends only to about 15 MeV, well below the threshold for
producing muons, this reaction is sensitive only to electron neutrinos.

\begin{figure}
\begin{minipage}[b]{0.5\linewidth}
\centering
\includegraphics[height=9.4cm]{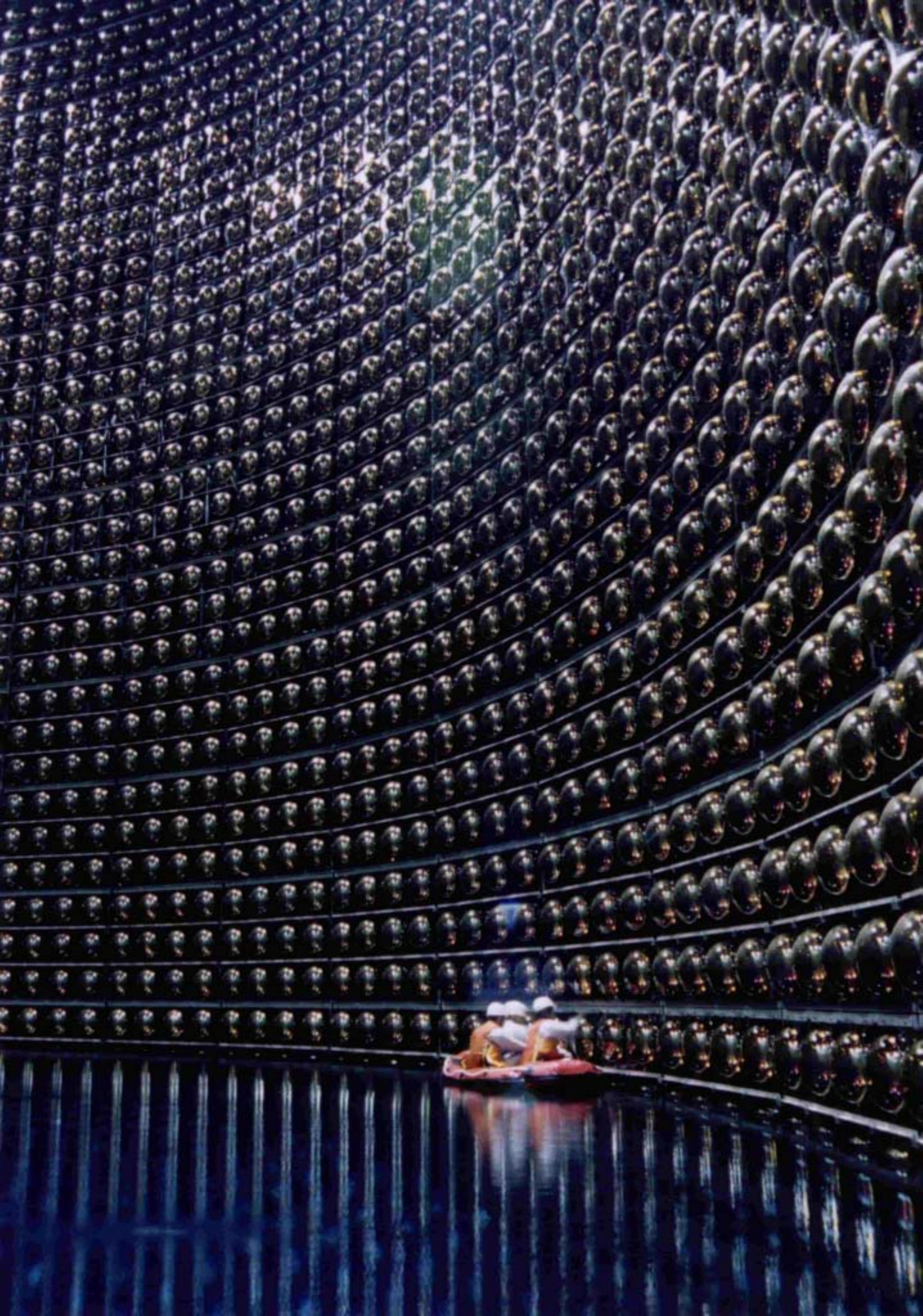}
\end{minipage}
\begin{minipage}[b]{0.5\linewidth}
\centering
\includegraphics[height=9.4cm]{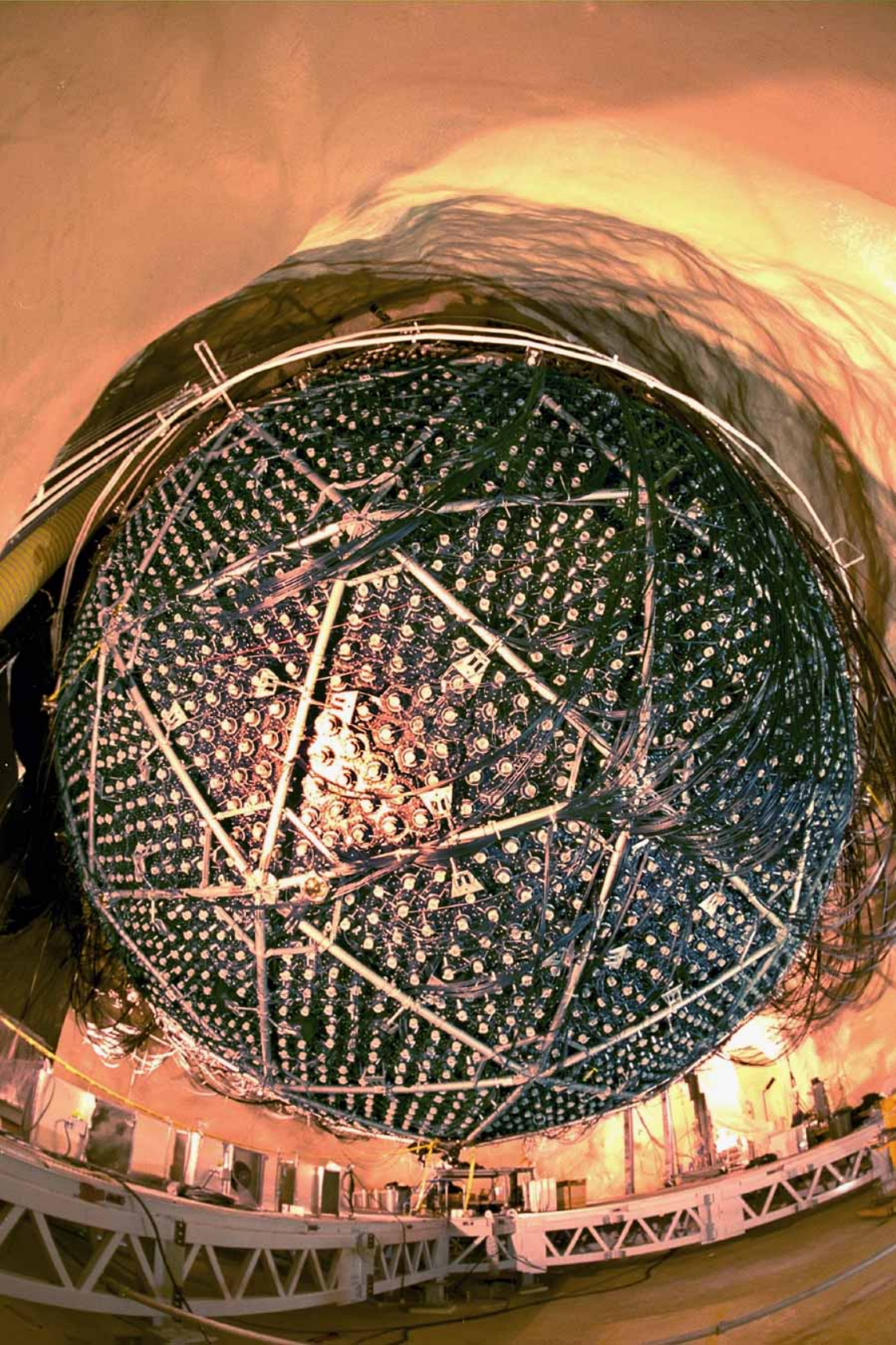}
\end{minipage}
\caption{The left panel shows the Super-Kamiokande detector during filling, with
scientists cleaning PMT surfaces as the water rises.  The right panel
is a fish-eye photo of the SNO detector and cavity, showing the PMTs
and support structure prior to cavity and detector filling.}
\label{fig:sksno}
\end{figure}

A second possible nuclear detection channel is neutral-current
scattering
\[ \nu_x + (N,Z) \rightarrow \nu_x^\prime + (N,Z)^*, \]
a process independent of the neutrino flavor.  If this scattering
leaves the nucleus in an excited state, the observable would be the
de-excitation of the nucleus, such as a decay $\gamma$ ray or the
breakup of the nucleus.  (An example will be given below, in the
discussion of SNO.)  Alternatively, neutrino elastic scattering 
(without nuclear excitation) is a coherent process at low energies,
with a cross section proportional to the square of the weak charge,
which is approximately the neutron number $N$ of the nucleus.
The signal then is the small recoil energy of the nucleus after
scattering.  

A third possibility is the scattering of neutrinos off electrons,
\[ \nu_x + e^- \rightarrow \nu_x^\prime + e^-,\]
with detection of the recoiling scattered electron.
Both electron- and heavy-flavor ($\nu_\mu$, $\nu_\tau$) solar neutrinos
can scatter off electrons, the former by charge and neutral currents,
and the latter by neutral currents only.  Consequently the cross
section for scattering heavy-flavor neutrinos is only about 0.15 that
for electron neutrinos.  This process provides a third way
of probing neutrino flavor, due to this differential
sensitivity.   An important aspect of electron-neutrino
scattering is its directionality:  for solar neutrino energies much above
the electron mass of 0.511 MeV, the electron scatters into a narrow
cone along the incident neutrino's direction.   This directionality 
provides a powerful tool for extracting solar neutrino events from 
background:  neutrino events correlate with the direction of the Sun, while
background events should be isotropic.  Thus neutrino events can be
identified as the excess seen at forward electron angles.

These various detection channels were exploited in two large-volume
water Cerenkov detectors that recorded events in real time and
provided flavor sensitivity.  

Super-Kamiokande (Fig.~\ref{fig:sksno}) is a detector consisting of 50 kilotons of ultra-pure water
within a cylindrical stainless steel tank, 39m in diameter and 42m tall.  
Two meters inside the walls a scaffold supports a dense array of 50-cm-diameter hemispherical 
photomultiplier tubes (PMTs), which face inward and view the inner 32 kilotons of
water.  Additional 20-cm tubes face outward, viewing the outer portion
of the detector that serves as a shield and as a veto.   A solar neutrino 
can interact in the inner detector, scattering off an electron.  The recoiling,
relativistic electron then produces a cone of Cerenkov radiation, a pattern
that can be reconstructed from the triggering of the phototubes that
surround the inner detector.  The detector is housed deep within Japan's
Kamioka Mine, approximately one kilometer underground.

The detector has produced three data sets.  In the first, 
obtained from 1996-2001, the detector was instrumented with an array of 11,146 
50-cm PMTs, corresponding to about 40\% coverage.    In November 2001 
one of the 50-cm PMTs imploded, creating
a powerful shock wave that propagated through the tank, destroying 60\%
of the phototubes.  The detector was subsequently rebuilt  with about
half the original number of phototubes, evenly spaced over the scaffold,
so that the coverage was reduced to 20\%.  The detector operated in this
SK-II phase in the years 2003-5.  In August 2006 the detector resumed
operations after a second reconstruction in which the phototube coverage
was restored to 40\% and other improvements made.  This marked the start of
the ongoing SK-III phase.  The SK-I and SK-II results \cite{SKII} for the high-energy
${}^8$B neutrino flux are $(2.35 \pm 0.02 \mathrm{(stat)} \pm 0.08 \mathrm{(sys)}) \times 10^6$/cm$^2$s and
$(2.38 \pm 0.05 \pm 0.16) \times 10^6$/cm$^2$/s, respectively, well below
the SSM predictions of Table \ref{table:one}.

The Sudbury Neutrino Observatory (SNO) (Fig.~\ref{fig:sksno}) was constructed at quite
extraordinary depth, two kilometers 
underground in Inco's Creighton nickel mine in Ontario, Canada.
The detector took data from May 1999 through November 2006, operating
in three different modes over its 7.5-year lifetime.   SNO employed a
one-kiloton target of heavy water, contained within a spherical acrylic vessel
six meters in radius.  This sphere was surrounded by an additional five meters
(seven kilotons) of very pure ordinary water, filling the rock cavity 
that housed the entire detector.  An
array of 9600 20-cm PMTs, mounted on a geodesic sphere surrounding the inner
vessel, provided 56\% coverage.  As in the case of Super-Kamiokande,
SNO operated with a threshold of about 5 MeV.  Thus it detected that
portion of the ${}^8$B solar neutrino spectrum from 5-15 MeV.

The choice of a heavy-water target allowed SNO experimentalists to 
exploit all three of the reaction channels described above, with their
varying flavor sensitivities
\begin{eqnarray}
\nu _x + e^- \rightarrow \nu_x^\prime +
e^{-}~~~&&\mathrm{ES:~elastic~scattering} \nonumber \\
\nu _e + d \rightarrow  p + p + e^{-}~~~&&\mathrm{CC:~charged~current} \nonumber \\
\nu _x + d \rightarrow  \nu_x^\prime + n +
p~~~&&\mathrm{NC:~neutral~current} \nonumber 
\end{eqnarray}
The elastic scattering (ES) reaction is the same as that employed by Super-Kamiokande, 
with its differing sensitivities to electron- and heavy-flavor neutrinos.  The charged-current (CC)
reaction on deuterium is sensitive only to electron-flavor neutrinos,
producing electrons that carry off most of 
the incident neutrino's energy (apart from the 1.44 MeV needed to
break a deuterium nucleus into p+p).   
Thus, from the energy distribution of the electrons, one can 
reconstruct the incident $\nu_e$ spectrum (and
possible distortions discussed below) more accurately than in the
case of ES.

The NC reaction, which is observed through the produced neutron,
provides no spectral information, but does measure the total solar
neutrino flux, independent of flavor. The SNO experiment has used
three techniques for measuring the neutrons. In the initial pure-D$_2$O 
phase the neutrons captured on deuterium,
producing 6.25 MeV $\gamma$s. In a second phase 2.7 tons of salt were
added to the heavy water so that Cl would be present to enhance
the capture, producing 8.6 MeV $\gamma$s. In both of these approaches
the NC and CC events can be
separated reasonably well because of the modest backward peaking
($\sim$ 1-$\cos{\theta}$/3) in the angular distribution of the
latter. This allowed the experimenters to determine the total and
electron neutrino fraction of the solar neutrino flux.  Finally, in the third
phase, direct neutron detection was provided in
pure D$_2$O by an array of ${}^3$He--filled proportional counters,
cleanly separating this signal from CC scattering.

SNO was constructed at very great depth and under clean-room 
conditions because of the need to suppress backgrounds.  In particular,
a minute amount of dust in the detector could have introduced 
environmental radioactivities that would have obscured the NC signal, a
single neutron.  The great advantage of the SNO detector was its
three distinct detection channels, sensitive to different
combinations of electron and heavy-flavor neutrinos.  Furthermore,  because the ES and
CC scattered electrons are measured in the same detector, several
important systematic effects cancel in the ratio of events.  The ES
reaction provided an important cross check on the consistency of
the results from the CC and NC channel.

\begin{figure}
\begin{center}
\includegraphics[width=12cm]{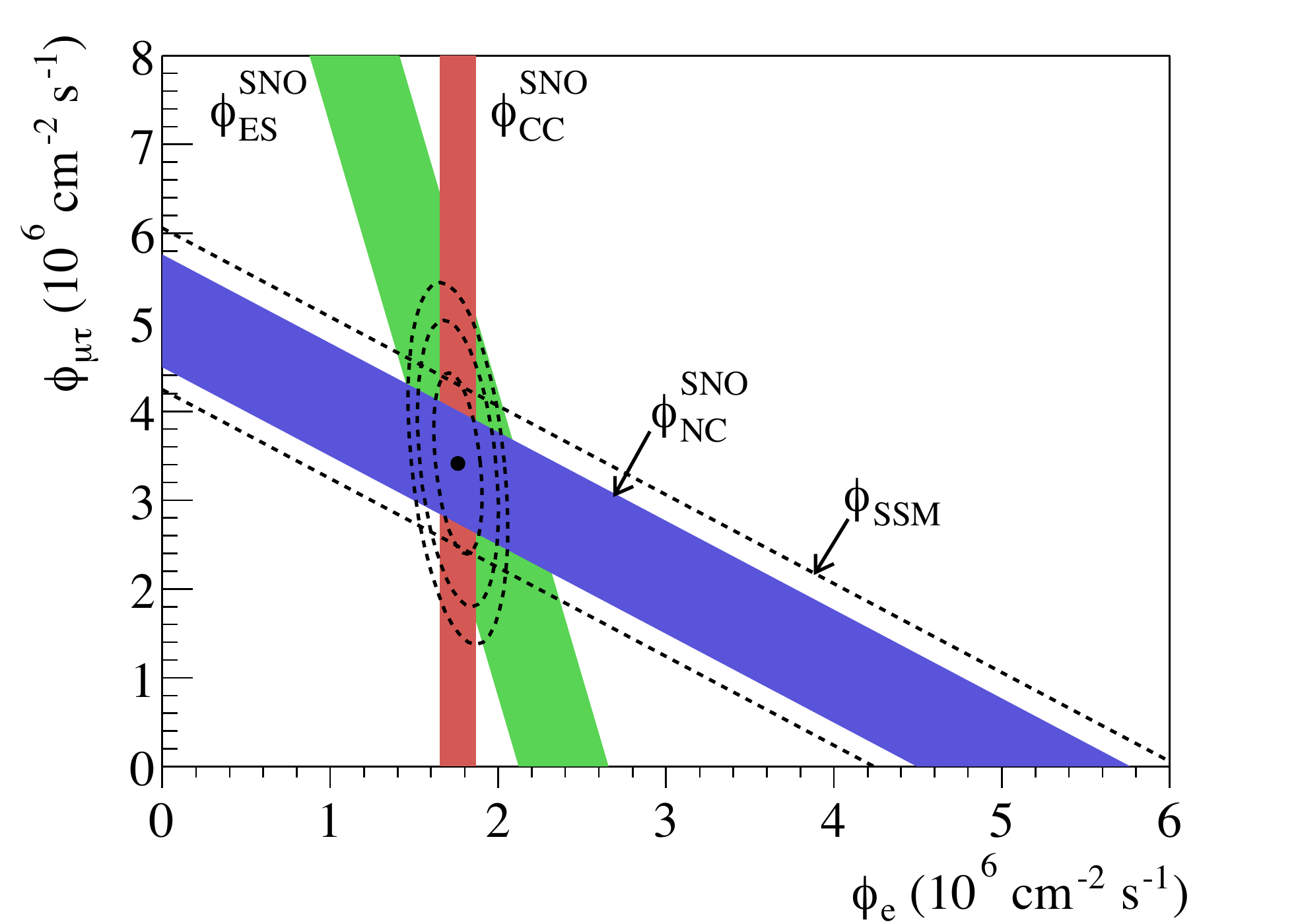}
\end{center}
\caption{ Results from the D$_2$O phase of the SNO experiment \cite{SNOsn}.  The allowed bands for
CC, NC, and ES reactions of solar neutrinos intersect to show a flux that is
one-third $\nu_e$s and two-thirds heavy-flavor
neutrinos.  There is agreement between the NC total-flux measurement and
the predictions of the SSM (band indicated by the dashed lines).  Figure courtesy of the
Sudbury Neutrino Observatory collaboration.}
\label{fig:SNOresults}
\end{figure}

The first  results \cite {SNOI} from SNO are shown in Fig.~\ref{fig:SNOresults}. 
The final SNO Phase I (neutron capture on deuterium) deduced fluxes are \cite{SNOsn}
\begin{eqnarray}
\phi_{CC} = \left[1.76^{+0.06}_{-0.05} \mathrm{(stat.)}^{+0.09}_{-0.09} \mathrm{(syst,)} \right] \times
10^6/\mathrm{cm}^2\mathrm{s} \nonumber \\
\phi_{ES} = \left[2.39^{+0.24}_{-0.23} \mathrm{(stat.)}^{+0.12}_{-0.12} \mathrm{(syst.)} \right] \times
10^6/\mathrm{cm}^2\mathrm{s} \nonumber \\
\phi _{NC} = \left[ 5.09^{+0.44}_{-0.43} \mathrm{(stat.)}^{+0.46}_{-0.43} \mathrm{(syst.)} \right]
\times 10^6/\mathrm{cm}^2\mathrm{s} \nonumber
\end{eqnarray}
Fig.~\ref{fig:SNOresults} shows that the total flux (NC) is in 
agreement with the SSM prediction, but the neutrino flavor has been changed:
about two-thirds of the electron
neutrinos produced in the Sun arrive on Earth as heavy-flavor
(muon or tauon) neutrinos.   The Davis detector and the
CC channel in SNO are blind to these heavy flavors, seeing only
the portion with electron flavor.  Thus the solar neutrino problem
was not a matter of missing neutrinos, but rather one of neutrinos
in hiding.  The implications of this discovery -- that neutrinos are
massive and violate flavor -- are profound, indicating that our
standard model of particle physics is incomplete.

\subsection{Neutrino Mass and Oscillations}
The phenomenon by which a massive neutrino of one flavor
changes into one of a second flavor is called neutrino oscillations.  
Neutrino oscillations have been shown to be responsible not
only for the missing solar neutrinos in Davis's experiment, but also
for the missing atmospheric neutrinos that will be discussed
in the next chapter.  Neutrino
oscillations can be altered by the presence of matter or the
presence of other neutrinos.  For this reason, astrophysical ``laboratories" for
studying neutrinos -- the Sun, supernovae, the early universe -- are
of great interest because of the unique conditions they
provide, including very
long ``baselines" over which neutrino propagate and enormous matter
densities and neutrino fluences.

Neutrino oscillations originate from two distinct sets of labels
carried by neutrinos.  One is flavor, a property of the 
weak interaction: an electron neutrino is defined as the neutrino
accompanying a positron in $\beta$ decay.  The other possible label
is mass.  If a neutrino
has a mass $m$, it propagates through free 
space with an energy and three momentum related by
$\omega = \sqrt{\vec{k}^2 + m^2}$.    Thus neutrino states
can be labeled according to flavor, and also labeled according
to their masses.

However nothing requires the neutrinos of definite flavor to
be coincident with the neutrinos of definite mass.  (In fact, in the
analogous case of the quarks, it has long been known that the
flavor (or weak interaction) eigenstates are not identical to the
mass eigenstates: That is, the up
quark decays not only to the down quark, but also occasionally
to the strange quark.)  Neutrino oscillations occur when
the mass eigenstates $|\nu_1 \rangle$ and $|\nu_2 \rangle$ (with masses
$m_1$ and $m_2$) are related to the weak interaction eigenstates by
\begin{eqnarray}
|\nu_e\rangle  &=& \cos \theta_v |\nu_1\rangle  + \sin \theta_v|\nu_2 \rangle  \nonumber \\
|\nu_\mu\rangle &=& - \sin \theta_v |\nu_1 \rangle + \cos \theta_v |\nu_2 
\rangle \nonumber
\end{eqnarray}
where $\theta_v$, the (vacuum) mixing angle, is nonzero.  (Here, for
simplicity, we consider just two neutrinos -- the generalization to three
flavors adds little new.)
  
\begin{figure}
\begin{center}
\includegraphics[width=12cm]{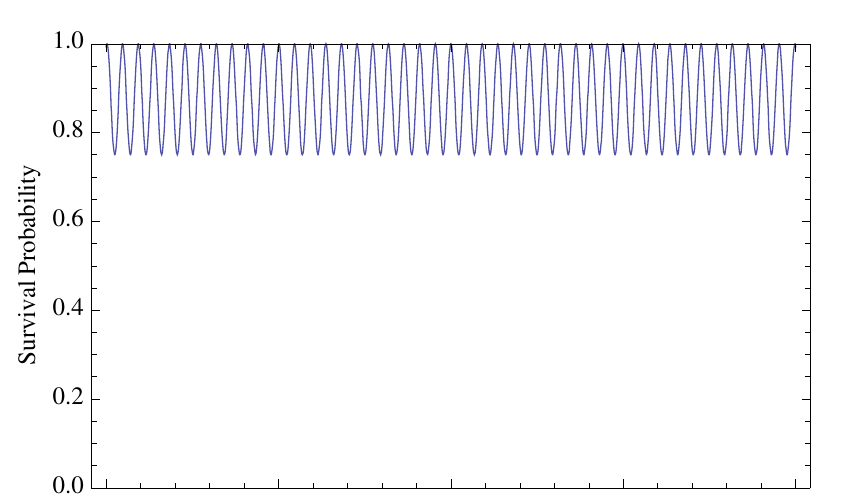}
\includegraphics[width=12cm]{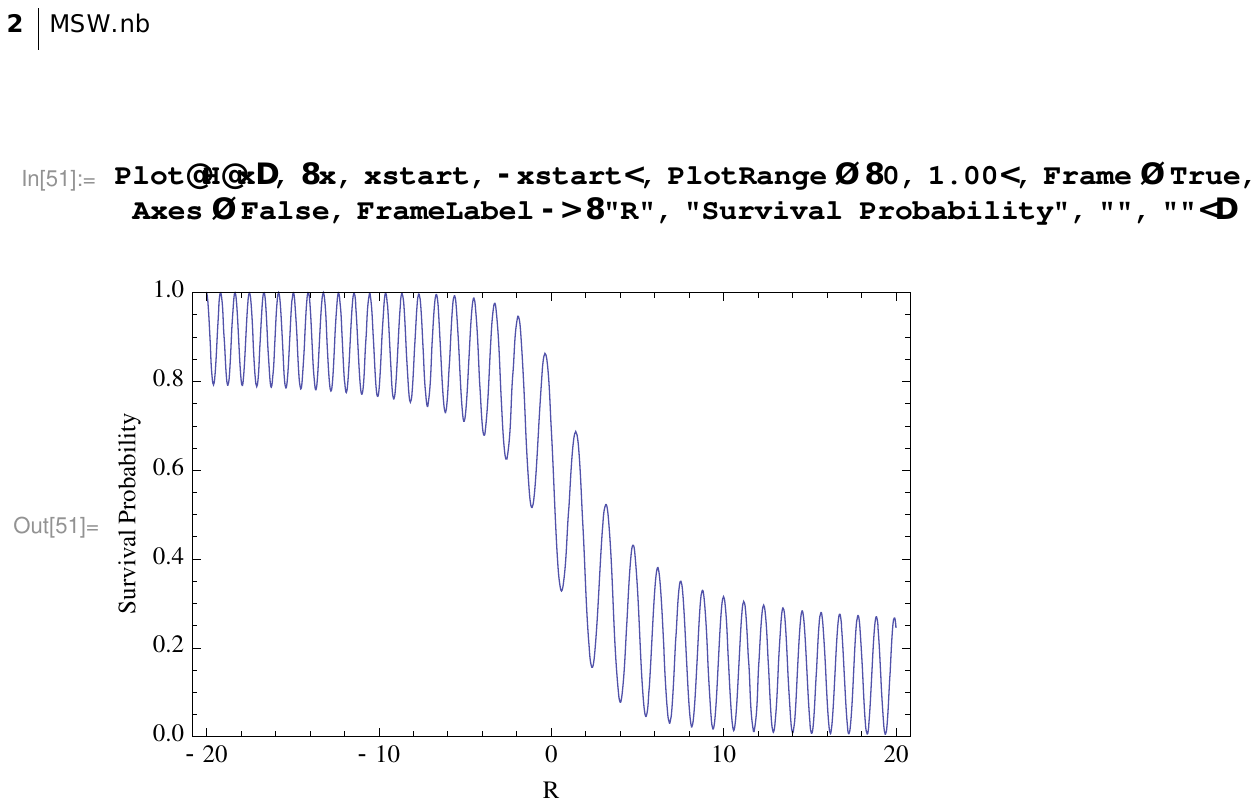}
\end{center}
\caption{A simple example illustrating the MSW mechanism.  The top frame
shows vacuum oscillations for a $\nu_e$ created at $R=-20$ and propagating
to the right, for $\theta_v$=15$^\circ$.  The average $\nu_e$ survival
probability is large, 87.5\%.  (Here
the distance R is given in units related to the oscillation length,
$4 E \cos{2 \theta}/(\delta m^2 \sin^2{2 \theta}$).)  In the bottom frame an electron
density $\rho(R)$ has been added proportional to $1- (2/\pi) \tan^{-1}{a R}$, with $a$ chosen to
guarantee adiabaticity, and normalized so that  1) $\rho(r) \rightarrow 0$ as $R \rightarrow \infty$;
2) the matter effects cancel the vacuum mass difference for $R \sim 0$ (the MSW crossing point);
and 3) the matter effects reverse the sign of $\delta m^2$ as $R \rightarrow -\infty$.  Thus these are the
MSW conditions described in the text.  A $\nu_e$ created 
at high density ($R =-20$), where it approximately coincides with the local heavy-mass
eigenstate, adiabatically propagates to low-density ($R=+20$), where it approximately
coincides with the $\nu_\mu$.   Thus the $\nu_e$
survival probability at $R=20$ is much reduced, compared to the vacuum case.  Note
that the local oscillation length is maximal near the crossing point.}
\label{fig:osc}
\end{figure}

In this case a state produced as a $|\nu_e\rangle$
or a $|\nu_\mu\rangle$ at some time $t$ --- for example, a neutrino
produced in $\beta$ decay in the Sun's core --- does not remain a pure flavor eigenstate
as it propagates away from the source.  The different
mass eigenstates comprising the neutrino will accumulate different
phases as the neutrino propagate downstream, a phenomenon known as 
vacuum oscillations (vacuum because the experiment is done in free
space).  While at time $t$=0 the neutrino is a flavor eigenstate
\[ |\nu(t=0)\rangle  = |\nu_e \rangle = \cos \theta_v |\nu_1\rangle  + \sin \theta_v|\nu_2 \rangle ,\]
the accumulate phases depend on the mass
\[e^{i(\vec{k} \cdot \vec{x} - \omega t)} =
e^{i [ \vec{k} \cdot \vec{x} - \sqrt{m_i^2 + k^2}~t ]} . \]
If the neutrino mass is small compared to the neutrino 
momentum/energy, one finds
\[ |\nu(t) \rangle = e^{i(\vec{k} \cdot \vec{x} - kt
-(m_1^2+m_2^2)t/4k)} \left( \cos \theta_v |\nu_1 \rangle e^{i \delta m^2 t/4k}
+ \sin \theta_v |\nu_2 \rangle e^{-i \delta m^2 t/4k} \right) \] 
There is a common average phase (which has no physical
consequence) as well as a beat phase that depends on
\[ \delta m^2 = m_2^2 - m_1^2.  \]
From this one can find the probability that 
the neutrino state remains a $|\nu_e\rangle$ at time t
\[ P_{\nu_e} (t) = | \langle \nu_e | \nu(t) \rangle |^2  =
1 - \sin^2 2 \theta_v \sin^2 \left({\delta m^2c^4 x\over 4 \hbar c E} \right)  \]
The probability oscillates from 1 to  $1-\sin^2 2 \theta_v$ and back to 1
over an oscillation length scale
\[ L_o = {4 \pi \hbar c E \over \delta m^2 c^4} ,\]
as depicted in Fig.~\ref{fig:osc}.
In the case of solar neutrinos, if $L_o$ were comparable to or shorter than one astronomical unit, a 
reduction in the solar $\nu_e$ flux would be expected in terrestrial detectors.
  
The suggestion that the solar neutrino problem could
be explained by neutrino oscillations was first made by
Pontecorvo in 1958, who pointed out the analogy with $K_0 \leftrightarrow \bar 
K_0$     
oscillations.  If the Earth-Sun separation is much larger than $L_o$, one expects 
an average flux reduction due to oscillations of
\[ 1 - {1 \over 2} \sin^2 2 \theta_v . \]
For a 1 MeV neutrino, this requires $\delta m^2c^4 \gg 10^{-12}$ eV$^2$.  
But such a reduction $-$ particularly given the initial theory prejudice that neutrino mixing
angles might be small $-$ did not seem sufficient
to account for the factor-of-three discrepancy that emerged from Davis's early
measurements.

The view of neutrino oscillations changed 
when Mikheyev and Smirnov \cite{MS} showed in 1985 that neutrino
oscillations occurring in matter -- rather than in vacuum -- could
produce greatly enhanced oscillation probabilities.   This
enhancement comes about because neutrinos propagating through matter 
acquire an additional mass due to their interactions with the
matter.  In particular, because the Sun contains many electrons, 
the electron neutrino becomes heavier in proportion to the
local density of electrons.   An enhanced probability for oscillations can
result when an electron neutrino passes from a high-density region
(such as the solar core) to a low-density one (such as the surface
of the Earth).  This matter enhancement is
called the MSW mechanism after
Mikheyev, Smirnov, and Wolfenstein \cite{Wolf} (who first described the 
phenomenon of neutrino effective masses).

To explain this enhancement, consider the case where the mixing
angle in vacuum, $\theta_v$, is small  and $m_2 > m_1$.
Then $|\nu_e\rangle  \sim  |\nu_1\rangle \equiv |\nu_L(\rho=0) \rangle$
where $\rho$ is the local electron density, that is,
the $\nu_e$ and the light vacuum eigenstate $|\nu_L(\rho=0) \rangle$ are almost identical.
(Correspondingly, the heavy eigenstate $|\nu_2 \rangle \equiv |\nu_H(\rho=0) \rangle
 \sim |\nu_\mu \rangle$ in vacuum.)
Now what happens in matter?  As matter makes the $\nu_e$
heavier in proportion to the electron density, if that density
is sufficiently high, clearly the electron neutrino
must become the (local) heavy mass eigenstate.  That is, 
$|\nu_e\rangle  \sim |\nu_H(\rho \rightarrow \infty) \rangle$ (and
consequently $|\nu_\mu\rangle  \sim  |\nu_L(\rho \rightarrow \infty) \rangle$).
That is, we conclude that there must be a local mixing angle $\theta(\rho)$
that rotates from $\theta_v \sim 0$
in vacuum to $\theta(\rho) \sim \pi/2$ as $\rho \rightarrow \infty$. 

MSW enhancement occurs when the density changes between
neutrino production and detection.    In particular, electron neutrinos
produced in the high-density solar core are created as heavy mass
eigenstates.  If these neutrinos now propagate to the solar surface
adiabatically  -- this means that changes in the solar density scale height
$1/\rho~ d \rho/dx$ are 
small over an oscillation length, at all points along
the neutrino trajectory -- they will remain on the heavy-mass trajectory,
and thus exit the Sun as $|\nu_H(\rho=0)\rangle =
|\nu_2 \rangle \sim |\nu_\mu \rangle$. That is, there
will be an almost complete conversion of the $\nu_e$s produced
in the solar core to $\nu_\mu$s.  The MSW mechanism is an example
of an avoided level crossing, a familiar phenomenon in quantum
mechanics.

A schematic comparison of vacuum and matter-enhanced oscillations
is shown in Fig.~\ref{fig:osc}.  The matter transition between electron
and muon flavors is centered around a density where the vacuum mass 
difference is just compensated by the matter contributions.

The results from SNO and Super-Kamiokande, from earlier solar
neutrino experiments, and from the
reactor experiment KamLAND, have determined the parameters
governing solar neutrino oscillations quite precisely \cite{SNOKam}.   Unlike the
example given above, the relevant mixing angle is rather large,
$\theta_{12} \sim 34^\circ$.   The vacuum mass difference,
$\delta m_{12}^2 \sim 7.94 \times 10^{-5}$ eV$^2$, leads to important
matter effects in the higher energy portion of the solar neutrino
spectrum, thus influencing the rates found in the SNO, Super-Kamiokande,
and chlorine experiments.  These effects produce a characteristic,
energy-dependent distortion of the solar $\nu_e$ spectrum.

\subsection{Solar Neutrinos: Outlook}
Neutrino oscillations proved to be responsible for both the solar
neutrino problem and the atmospheric neutrino problem discussed
in the next chapter.
This phenomenon requires both flavor mixing and neutrino mass,
phenomena that can be accommodated in various extensions of the
standard model of particle physics.  The great interest in neutrino astrophysics
stems in part from the expectation that newly discovered neutrino properties
may help us formulate the correct extension.  Indeed, the tiny neutrino
mass differences deduced from the solar and atmospheric problems are
compatible with theories where the neutrino mass is inversely 
proportional to a scale for new physics of about 10$^{15}$ GeV.  This is
close to the Grand Unified scale where supersymmetric extensions
of the standard model predict that the strengths of the fundamental forces
unify.

\begin{figure}
\begin{center}
\includegraphics[width=12cm]{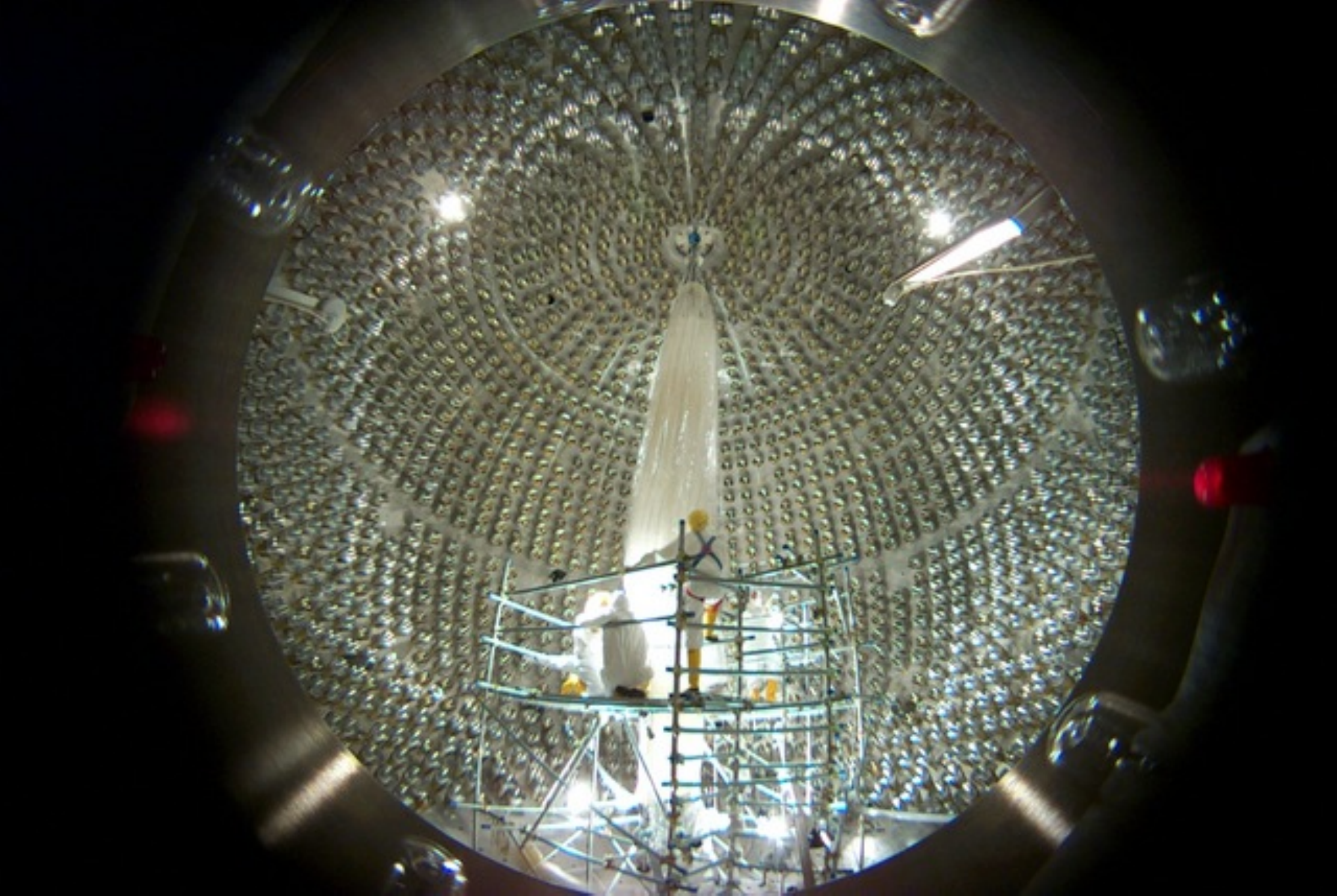}
\end{center}
\caption{ The Borexino inner vessel during installation.}
\label{fig:Borexino}
\end{figure}

There remain important, unresolved issues in solar neutrino physics.   The
first direct (real-time) measurements of the dominant, low-energy neutrino fluxes -- the pp and ${}^7$Be
neutrinos -- are just beginning.  Borexino \cite{Borexino}, a scintillation detector operating in the Italy's
Gran Sasso laboratory (Fig.~\ref{fig:Borexino}), began taking data on the ${}^7$Be flux in May 2007.  
One goal is to test the prediction that the survival probability of ${}^7$Be $\nu_e$s
will be larger than that for the higher energy neutrinos, since matter effects 
are more significant for the latter.

There are also interesting developments involving the SSM.
One of the important validations of the SSM has come through helioseismology,
the measurement of solar surface fluctuations, as deduced from the Doppler
shifts of spectral lines.  The observed patterns can be inverted to determine
properties of interior acoustic and gravity waves, including the sound speed
as a function of the solar radius.  For about a decade the agreement between SSM
predictions and observation had been excellent.  

But recently improved three-dimensional models of photospheric absorption lines
have led to a 30\% downward revision in convective-zone metal abundances 
(that is, abundances of elements heavier than helium).  As
discussed earlier, the SSM fixes the Sun's zero-age metallicity to today's surface
abundances.  If the new abundances are employed in the SSM, the resulting
changes in the opacity alter
both neutrino flux predictions and the sound speed profile.  The
discrepancy between the SSM sound speed profile and helioseismology is particularly
significant in the Sun's upper radiative zone.
If this difficulty persists, some of the
assumptions of the SSM, such as a homogeneous zero-age Sun, may have to be
critically re-examined.

\section{Atmospheric Neutrinos}
\noindent
The atmospheric neutrino problem developed very much in parallel with
the solar neutrino problem and also involved missing neutrinos.   The first
definitive claim that neutrinos are massive came the atmospheric neutrino group
associated with Super-Kamiokande, in 1998.   The oscillations seen in atmospheric
neutrinos differ from those seen in solar neutrinos, resulting from the coupling
of a different pair of neutrinos.

\subsection{The Neutrino Source}When primary cosmic-ray protons and nuclei hit the upper atmosphere, the ensuing
nuclear reactions with atmospheric oxygen and nitrogen nuclei produce secondaries
such as pions, kaons, and muons.    Atmospheric neutrinos arise from the decay of 
these secondaries.   For energies less
than $\sim$ 1 GeV, the secondaries decay prior to reaching the Earth's surface
\begin{eqnarray}
\pi^{\pm} (K^{\pm}) &\rightarrow &\mu^{\pm} + \nu_{\mu}
(\overline{\nu}_{\mu}), \nonumber\\ \mu^{\pm} &\rightarrow & e^{\pm} +
\nu_e (\overline{\nu}_e) +  \overline{\nu}_{\mu} (\nu_{\mu}).
\end{eqnarray}
Consequently one expects the ratio
\begin{equation}
r = (\nu_e + \overline{\nu}_e) / (\nu_{\mu} + \overline{\nu}_{\mu})
\end{equation}
to be approximately 0.5 in this energy range. Detailed Monte Carlo
calculations, including the effects of muon
polarization, give $  r \sim 0.45$.  This ratio should be rather insensitive to
theoretical uncertainties.   It does not depend on absolute fluxes, and as
a ratio of related processes, one expects many sources of systematic error 
to cancel.  Indeed, while various groups have estimated  this ratio, sometimes
starting with neutrino fluxes which vary in magnitude by up to
25\%, agreement in the ratio has been found at the level of a few percent. 
This agreement persists at higher energies, where $r$ decreases because
higher energy muons survive passage through the atmosphere, due to the
effects of time dilation.

Atmospheric neutrinos are a very attractive astrophysical source for 
experimenters.  Apart from relatively minor geomagnetic effects, atmospheric 
neutrino production is uniform over the Earth.  Thus an experimenter, operating
an underground detector at some location, can make use of a set of nearly equivalent
neutrino sources at distances ranging from 10s of kilometers (directly overhead) to 13,000
kilometers (directly below, produced on the opposite side of the Earth).
Effects such as neutrino oscillations, which depend on the distance from the source 
to the target, might show up as a characteristic dependence of neutrino flux on
the zenith angle, provided the relevant
oscillation length is comparable to or less than the Earth's diameter.
(Note that the solar neutrino $\delta m_{12}^2$, for atmospheric
neutrinos of energy $\sim$ 1 GeV, would not satisfy this condition, as the
oscillation length is several times the Earth's diameter.)

\subsection{Atmospheric Neutrinos and Proton Decay Detectors}
The atmospheric neutrino anomaly grew out of efforts to build large underground detectors for
proton decay, one of the phenomena expected in the Grand Unified Theories that
were formulated in the late 1970s and early 1980s.  As atmospheric neutrinos and proton
decay would deposit very similar energies in such detectors, studies of atmospheric
neutrinos were a natural second use of such detectors.   Significant indications
of an anomaly came from the IMB \cite{IMB} and Kamiokande \cite{Kamioka} proton decay detectors.  IMB first noticed a possible deficit of neutrino-induced muon events in 1986, while
Kamiokande established a deficit in excess of 4$\sigma$ by 1988.   By 1998
this anomaly was also apparent in data from the Soudan detector and from Super-Kamiokande.

The quantity determined in such experiments is a ratio 
(observed to predicted) of ratios
\[ R = {(\nu_{\mu} / \nu_e)_{\rm data} \over (\nu_{\mu} / \nu_e)_{\rm
Monte  Carlo} } \]
where the numerator is determined experimentally, and the denominator
calculated.   Agreement between data and theory thus requires $R \sim1$.
Early experimenters faced a difficulty in evaluating this ratio
due to limited statistics:  the counting rates were too low to allow a detailed 
analysis based on the zenith angle, that is, based on the neutrino path length.
This changed with the construction of Super-Kamiokande, which provided a fiducial
volume of about 20 ktons.  An early analysis from Super-Kamiokande found
\[  R=0.61 \pm 0.03 ({\rm stat}) \pm 0.05 ({\rm syst}) \]
for sub-GeV events which were fully contained in the detector and
\[ R=0.66 \pm 0.05 ({\rm stat}) \pm 0.08 ({\rm syst}) \]
for fully- and partially-contained multi-GeV events.   In addition, the 
collaboration presented an analysis in 1998, based on 33 kton-years of data,
showing a zenith angle dependence inconsistent with theoretical calculations
of the atmospheric flux, in the absence of oscillations \cite{SKatmos}.  This indicated
a distance dependence in the muon deficit, a signature of oscillations.  
Furthermore the parameters of the oscillation, especially $5 \times 10^{-4}$ eV$^2 <
\delta m_{23}^2 < 6 \times 10^{-3}$ eV$^2$, differed from those that would
later be determined from solar neutrinos.  The collaboration concluded
that the data were consistent with the two-flavor oscillation $\nu_\mu \rightarrow
\nu_\tau$.  This was the first definitive claim for massive neutrinos.

\begin{figure}
\begin{center}
\includegraphics[width=12cm]{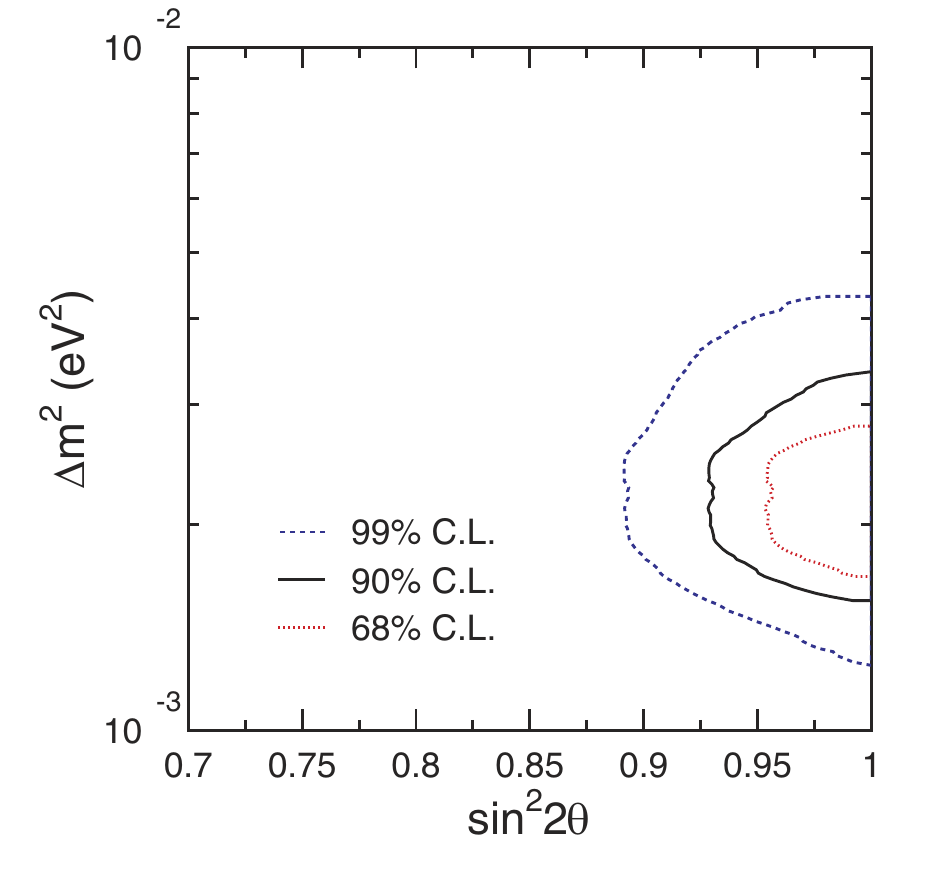}
\end{center}
\caption{ The Super-Kamiokande I analysis of atmospheric neutrino oscillations \cite{SKatmos}.}
\label{fig:atmospheric}
\end{figure}

SK-I collected approximately 15,000 atmospheric neutrino events
in nearly five years of running.
The collaboration's zenith-angle analysis of the data found
evidence of a first oscillation minimum at $L/E \sim 500$ km/GeV, so that
$L_o \sim$ 1000 km for a 1 GeV muon neutrino.   The full analysis of
oscillation parameters (see Fig.~\ref{fig:atmospheric}) gives a best-fit
$\delta m_{23}^2$ of $2.1 \times 10^{-3}$, a value clearly distinct from the
solar neutrino mass difference.  Most intriguing, the mixing angle appears to be
very close to 45$^\circ$ -- equal mixtures of two mass eigenstates.  

\subsection{Outlook}
\begin{figure}
\begin{center}
\includegraphics[width=12cm]{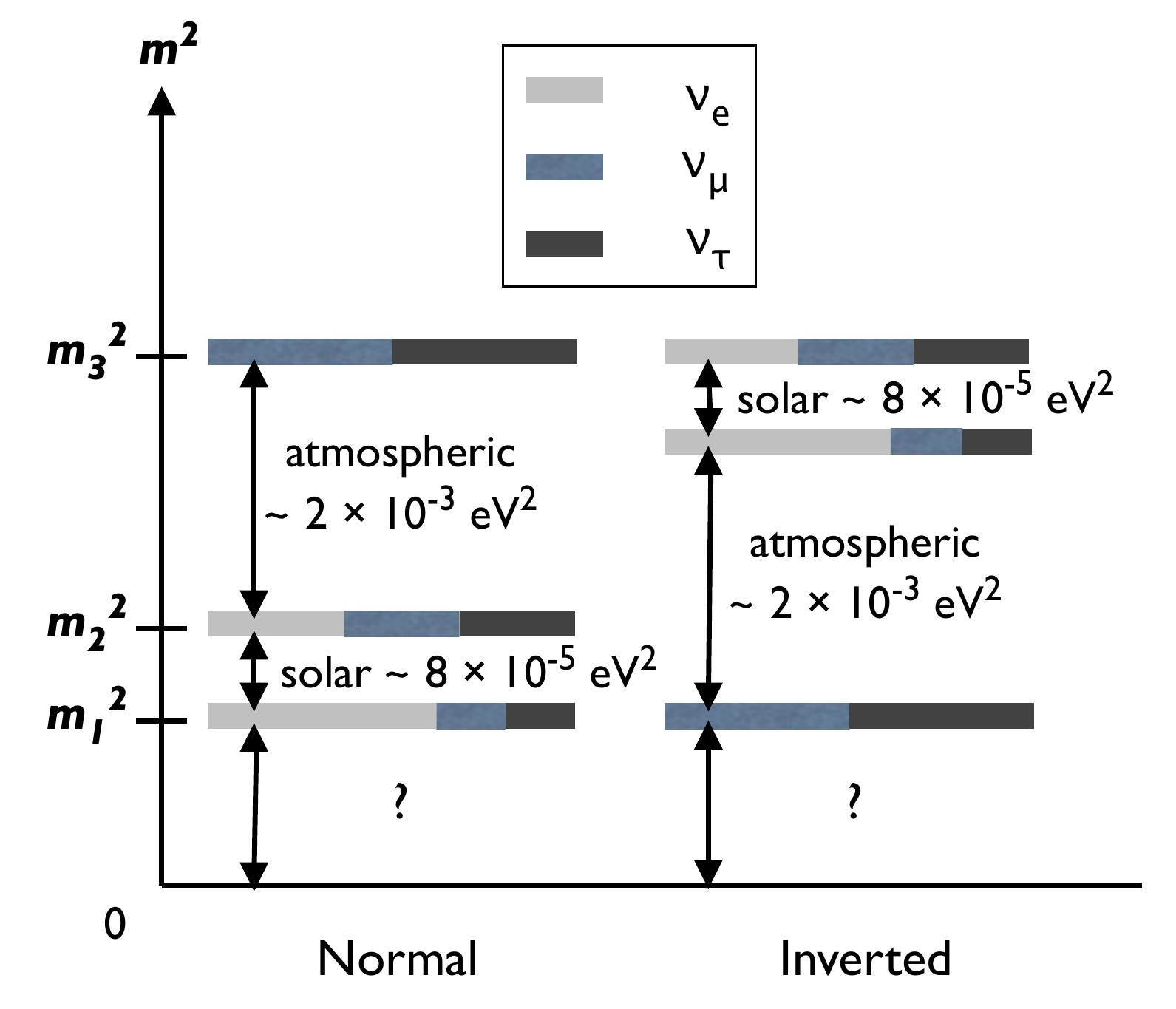}
\end{center}
\caption{ Illustration of the two level schemes that are possible for the light
neutrinos, given that no matter effects have yet been seen in atmospheric neutrinos.}
\label{fig:hierarchy}
\end{figure}

While a great deal of new physics has been learned from experiments on atmospheric
and solar neutrinos, several important questions remain \cite{APS}:
\begin{itemize}
\item Oscillation experiments are sensitive to differences in the squared masses.  They are
not sensitive to absolute neutrino masses.  We do know, from the atmospheric $\delta m^2$,
that at least one neutrino must have a mass $\gsim$ 0.04 eV.  But the only laboratory bound,
from tritium $\beta$ decay experiments, would allow neutrino masses 50 times greater.  That
is, the three light neutrinos might be nearly equal in mass, split by the tiny mass differences
indicated by solar and atmospheric neutrino oscillations.
\item Matter effects (from passage through the Earth) have not been seen in atmospheric
neutrino experiments.  This leaves open two possible orderings of the mass eigenstates, as
illustrated in Fig.~\ref{fig:hierarchy}. 
\item The solar and atmospheric mixing angles $\theta_{12}$ and $\theta_{23}$ have been determined,
but a third mixing angle, $\theta_{13}$, is so far only bounded by the results from reactor 
experiments, $\sin{\theta_{13}} \lsim 0.17$.
\item There are three CP-violating phases in the matrix that describes the relationship
between neutrino mass and flavor eigenstates, one of which could be detected
by looking for differences between certain conjugate oscillation channels, such as
$P(\nu_e \rightarrow \nu_\mu)$ and $P(\bar{\nu}_\mu \rightarrow \bar{\nu}_e)$.  Finding
such a difference is important to theories that attribute the excess of
matter over antimatter in our universe to leptonic CP violation.
\item The neutrino, lacking an electric or any other charge that must flip sign under particle-antiparticle
conjugation, is unique among standard model particles in that it may be its own antiparticle.  
So far no measurement has been made that can distinguish this possibility (a Majorana
neutrino) from the case where the $\nu$ and $\bar{\nu}$ are distinct (a Dirac neutrino).  
Next-generation neutrinoless double $\beta$ decay experiments
\[ (N,Z) \rightarrow (N-2,Z+2) + 2 e^- \]
could settle this issue, however.   This process requires lepton number violation and 
Majorana masses.  The two remaining CP-violating phases are Majorana phases
that can affect neutrinoless double $\beta$ decay rates.
\item No compelling argument has been given to account for the large mixing angles deduced
from atmospheric and solar neutrino oscillations.  These angles differ markedly from their
measured counterparts among the quarks. The special value of
the atmospheric angle $\theta_{23} \sim 45^\circ$
is particularly curious.
\end{itemize}
While some of these questions may be answered in terrestrial experiments, neutrino
astrophysics will continue to offer unique environments for probing fundamental neutrino
properties.  Several examples are given in the chapter on neutrino cooling.

\section{Supernovae Neutrinos and Nucleosynthesis}
The bursts associated with a core collapse supernova are among
the most interesting sources of neutrinos in astrophysics \cite{models}.  A massive
star, in excess of 10 solar masses, begins its lifetime burning
the hydrogen in its core under the conditions of hydrostatic
equilibrium.  When the hydrogen is exhausted, the core contracts
until the density and temperature are reached where 3$\alpha \rightarrow
^{12}$C can take place.  The helium is then burned to exhaustion.
This pattern (fuel exhaustion, contraction, heating, and ignition of the 
ashes of the previous burning cycle) repeats several times,
leading finally to the explosive burning of Si to Fe.
For a heavy star, the evolution is rapid due to the amount of energy
the star must produce to support itself against its own gravity.
A 25-solar-mass star would go through
the set of burning cycles in about 7 My, with the final explosive Si 
burning stage taking a few days.  The result is an 
``onion skin" structure of the precollapse star 
in which the star's history can be read by looking at the 
surface inward: there are concentric shells dominated by H, ${}^4$He,
${}^{12}$C, ${}^{16}$O and ${}^{20}$Ne, ${}^{28}$Si, and ${}^{56}$Fe
at the center. 

\subsection{The Explosion Mechanism and Neutrino Burst}
The source of energy for this evolution is nuclear binding energy.
A plot of the nuclear binding energy $\delta$ as a function of nuclear
mass shows that the minimum is achieved at Fe.  In a scale
where the ${}^{12}$C mass is picked as zero:
\begin{center}
${}^{12}$C~~~~~$\delta$/nucleon = 0.000 MeV \\
${}^{16}$O~~~~~$\delta$/nucleon = -0.296 MeV \\
${}^{28}$Si~~~~$\delta$/nucleon = -0.768 MeV \\
${}^{40}$Ca~~~~$\delta$/nucleon = -0.871 MeV \\
${}^{56}$Fe~~~~$\delta$/nucleon = -1.082 MeV \\
${}^{72}$Ge~~~~$\delta$/nucleon = -1.008 MeV \\
${}^{98}$Mo~~~~$\delta$/nucleon = -0.899 Mev
\end{center}
Once the Si burns to produce Fe, there is no further source
of nuclear energy adequate to support the star.  So as the last
remnants of nuclear burning take place, the core is largely
supported by degeneracy pressure, with the energy generation rate
in the core being less than the stellar luminosity.  The core
density is about 2 $\times 10^9$ g/cc and the temperature is
kT $\sim$ 0.5 MeV. 

Thus the collapse that begins with the end of Si burning is
not halted by a new burning stage, but continues.  As gravity
does work on the matter, the collapse leads to a rapid heating
and compression of the matter.  Sufficient heating of the Fe can release $\alpha$s and a few
nucleons, which are bound by $\sim$ 8 MeV.  At the same time, the electron chemical potential is
increasing.  This makes electron capture on nuclei and any free
protons favorable,
\[ e^- + p \rightarrow \nu_e + n. \]
As the chemical equilibrium condition is
\[ \mu_e + \mu_p = \mu_n + \langle E_\nu \rangle, \]
the increase in
the electron Fermi surface with density will lead to
increased neutronization of the matter, as long as neutrinos
freely escape the star.  These escaping neutrinos carry
off energy and lepton number.  Both the electron capture and
the nuclear excitation and disassociation take energy out of the electron gas,
which is the star's only source of support.  Consequently
the collapse is very rapid, with numerical simulations finding that 
the star's iron core ($\sim$ 1.2-1.5 solar mases) collapses
at about 0.6 of the free-fall velocity.

While the $\nu_e$s readily escape in the early stages of infall,
conditions change once the 
density reaches $\sim$ 10$^{12}$g/cm$^3$. 
At this point the neutrino scattering off the matter through
both charged current and coherent neutral current processes
begins to alter the transport.  The
neutral current neutrino scattering off nuclei is particularly
important, as the scattering amplitude is proportional to the total nuclear
weak charge, which is approximately the 
neutron number.  Elastic scattering transfers very little energy because
the mass of the nucleus is so much greater than the
typical energy of the neutrinos.  But momentum is exchanged. 
Because of repeated scattering the neutrino ``random walks" out of the star.  When the
neutrino mean free path becomes sufficiently short, the time
required for the neutrino to diffuse out of the high-density core
begins to exceed the time scale for the
collapse to be completed.  Above densities of about
10$^{12}$ g/cm$^3$, or $\sim$ 1\% of nuclear density, such neutrino trapping occurs.
Consequently, once this critical density is exceeded, 
the energy released by further gravitational
collapse is trapped within the star until after core bounce.  
Similarly, the star's lepton-number losses due to neutrino emission
largely cease.

For a neutron star of 1.4 solar masses and a radius of
10 km, an estimate of its binding energy is
\begin{equation}
 {G M^2 \over 2R} \sim 2.5 \times 10^{53} \mathrm{ergs}. 
\end{equation}
Thus this is roughly the trapped energy that will later be radiated in neutrinos,
after core bounce, as the proto-neutron star formed in the collapse cools. 

The collapse produces a shock wave that is critical to
subsequent ejection of the star's mantle.
The velocity of sound in matter rises with increasing density.
Late in the collapse the sound velocity in the inner portion of the iron core, 
with $M_{HC} \sim 0.6-0.9$ solar masses,
exceeds the infall velocity.  Any pressure
variations that may develop during infall can 
even out before the collapse is completed.  Consequently this
portion of the iron core collapses as a unit, retaining its density
profile.  

The collapse of the core continues until nuclear
densities are reached.  As nuclear matter is rather incompressible ($\sim$ 200 MeV/f$^3$),
the nuclear equation of state is effective in halting the collapse:
maximum densities of 3-4 times nuclear are reached, e.g.,
perhaps $6 \times 10^{14}$ g/cm$^3$.  The innermost shell of matter
reaches this supernuclear density first, rebounds, sending a 
pressure wave out through the inner core.  This wave
travels faster than the infalling matter, as the inner iron 
core is characterized by a sound speed in excess of the infall
speed.  Subsequent shells follow.  The resulting pressure
waves collect at the edge of the inner iron core -- the radius at which
the infall velocity and sound speed are equal.  
As this point reaches nuclear density and comes to
rest, a shock wave breaks out and begins its traversal of the 
outer core. 

Initially the shock wave may carry an order of magnitude more energy
than is needed to eject the mantle of the star (less than 10$^{51}$
ergs).  But as the shock wave travels through the outer iron core,
it heats and melts the iron that crosses the shock front, at a 
loss of $\sim$ 8 MeV/nucleon.  Additional energy is lost by neutrino
emission, which increases after the melting.   These losses are comparable to 
the initial energy carried by the shock wave.  Most simplified
(e.g., one dimensional) numerical models fail to produce a successful ``prompt"
hydrodynamic explosion, for this reason.   The shock stalls near the
edge of the iron core, instead of propagating into the mantle.

Most of the theoretical attention in the past decade has focused on the role
of neutrinos in reviving this shock wave, a process that becomes more 
effective in multi-dimensional models that account for convection.  In this delayed mechanism,
the shock wave stalls at a radius of 200-300 km, some tens of milliseconds
after core bounce.  But neutrinos diffusing out of the proto-neutron star
react frequently in the nucleon gas left in the wake of the shock wave,
depositing significant energy.  Over $\sim$ 0.5 seconds the increasing pressure
due to neutrino heating of this nucleon gas helps
push the shock outward.  This description is over-simplified -- a variety of
contributing effects are emerging from numerical simulations -- but there is
wide agreement that energy deposition by neutrinos is an essential ingredient
for successful explosions.

Regardless of explosion details, neutrinos dominate supernova energetics.
The kinetic energy of the explosion and
supernova's optical display account for less than 1\% of the available energy.
The remaining 99\% 
of the 3 $\times 10^{53}$ ergs released in the collapse is 
radiated in neutrinos of all flavors.  The timescale over 
which the trapped neutrinos leak out of the proto-neutron star
is about three seconds. 
The energy is roughly equipartitioned
among the flavors (a consequence of reactions
among trapped neutrinos that equilibrate flavor).  The detailed
decoupling of the emitted neutrinos from the matter -- which occurs at a
density of about $10^{11}-10^{12}$ g/cm$^3$ --does depend on flavor.
This leads to differences in neutrino temperatures, with electron
neutrinos being somewhat cooler ($T \sim$ 3.5 MeV) than the 
heavy-flavor neutrinos ($T \sim$ 6 MeV).
The radius for neutrino-matter decoupling defines a ``neutrinosphere" deep within the star, 
analogous to the familiar photosphere for optical emissions.

\begin{figure}
\begin{center}
\includegraphics[width=12cm]{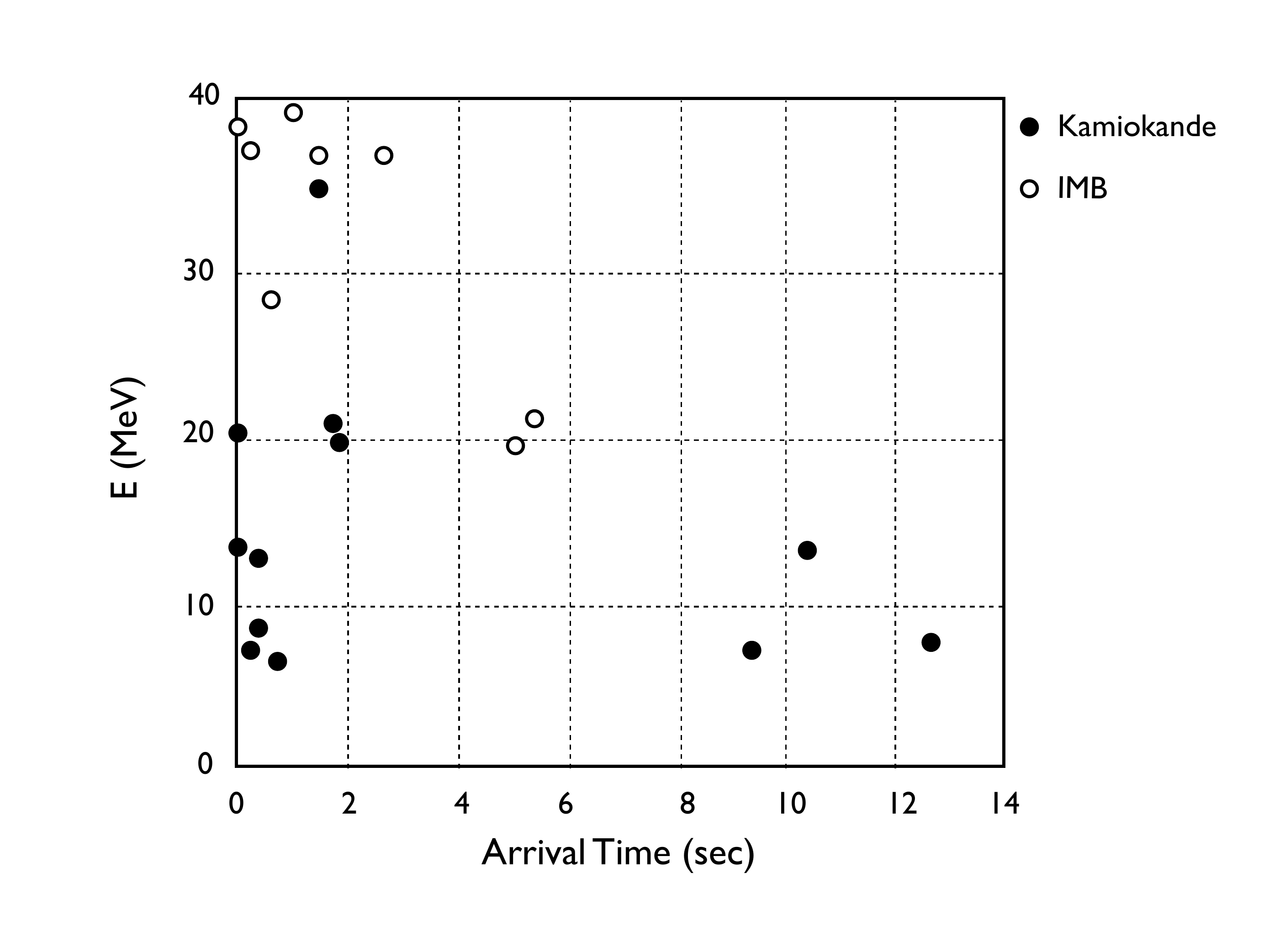}
\end{center}
\caption{ The timing and energy of neutrino events from SN1987A, as
observed in the Kamiokande and IMB detectors \cite{snnus}.}
\label{fig:SN1987Anew}
\end{figure}

The burst of neutrinos produced in a galactic core-collapse supernova
is detectable with instruments like Super-Kamiokande and SNO.   On February 23,
1987, a neutrino burst from a supernova in the Large Magellanic Cloud was
observed in the proton-decay detectors Kamiokande and IMB \cite{snnus}.   The optical
counterpart reached an apparent magnitude of about 3, and could be
observed easily in the night sky with the naked eye.  This supernova originated
160,000 light years from Earth.  Approximately 20 events were seen in the 
Kamiokande and IBM detectors, spread over approximately 10 seconds.  
Within the limited statistics possible with these first-generation detectors, the
number of events and the burst duration were consistent with standard 
estimates of the energy release and cooling time of a supernova.  The
neutrino data from the two detectors are shown in Fig.~\ref{fig:SN1987Anew}.

\begin{figure}
\begin{center}
\includegraphics[width=12cm]{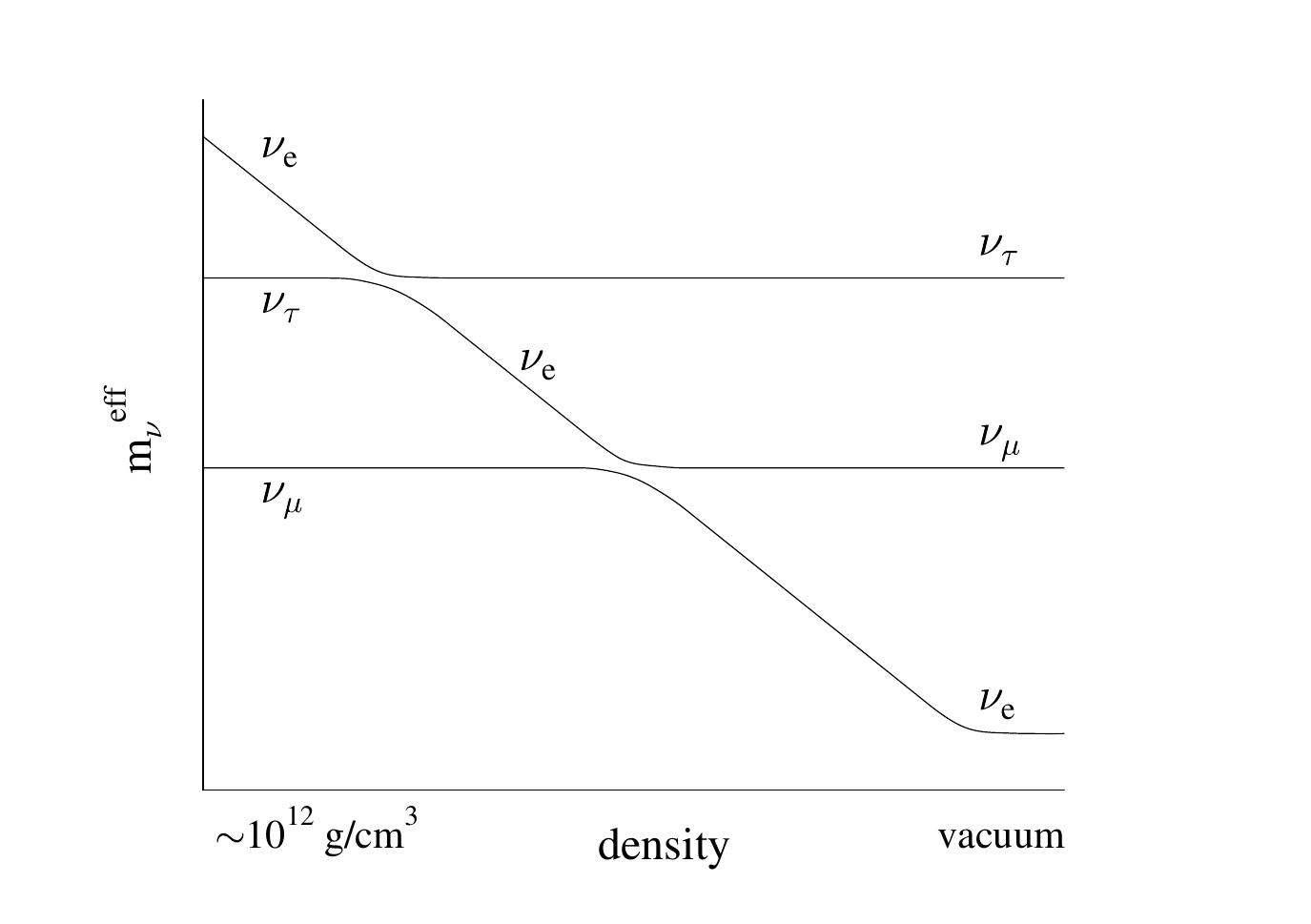}
\end{center}
\caption{A schematic illustration of the two level crossings that the $\nu_e$ would
experience, given the higher densities available in the mantle of a massive
star undergoing core collapse.}
\label{fig:threelevel}
\end{figure}

Temperature differences between neutrino flavors are interesting
because of oscillations and nucleosynthesis.  The discussion of matter
effects in the solar neutrino problem was limited to two flavors.  But the
higher densities found in core-collapse supernovae make all
three flavors relevant.  The three-flavor MSW
level-crossing diagram is shown in Fig.~\ref{fig:threelevel}.  One sees, in
addition to the neutrino ``level crossing"  $\nu_\mu \leftrightarrow \nu_e$
important for solar neutrinos, a second crossing of the $\nu_e$ with the
$\nu_\tau$.  The higher density characterizing this second crossing,
$\sim 10^4$ g/cm$^3$, is determined by atmospheric mass difference $\delta m_{23}^2$
and by the typical energy of supernova neutrinos, $\sim 10-20$ MeV.
This density is beyond that available in the Sun ($\lsim 10^2$ g/cm$^3$),
but far less than that of a supernova's neutrinosphere.  Consequently
this second level crossing alters neutrino flavor only after the supernova
neutrinos are free-streaming out of the star, with well defined spectra
that are approximately thermal.   This level crossing can exchange the
flavor labels on the $\nu_e$ and $\nu_\tau$ spectra, so that the $\nu_e$s
become hotter than the heavy-flavor neutrinos.  One would expect such 
an inversion to be apparent in terrestrial supernova neutrino detectors.

In fact this description oversimplifies the neutrino physics of supernovae.
The enormous neutrino densities encountered in a supernova lead to a new 
aspect of the MSW effect -- oscillations altered not by neutrino-electron
scattering, but by neutrino-neutrino scattering \cite{fuller}.   While the precise consequences of
this neutrino-neutrino MSW potential are still being debated, the effects reach
much deeper into the star and alter the flavor physics in distinctive ways.  
Supernovae likely provide the only environment in nature where
neutrino-neutrino interactions dominate the MSW potential.

\subsection{Supernova Neutrino Physics}
This novel neutrino-neutrino MSW potential is one of many reasons core-collapse supernovae
play an important role in neutrino astrophysics.  Others include:
\begin{itemize} 
\item The second level crossing involves the third, as yet unmeasured
mixing angle $\theta_{13}$.  This level-crossing occurs in the star's mantle and remains adiabatic $-$
the condition for flavor inversion $-$ for mixing angles $\sin^2{2 \theta_{13}} \gsim 10^{-4}$.
This kind of sensitivity to small mixing angles is not achievable in the next generation
of terrestrial neutrino experiments, which have goals of $\gsim 10^{-2}$.  Thus supernovae
may provide our only near-term hope of constraining this mixing angle, should it 
prove small.
\item Neutrinos from galactic supernovae will not be obscured by intervening matter or
dust, unlike optical signals.  Thus supernova neutrino bursts should, over the next few
hundred years, provide our most reliable measure of the contemporary rate of
galactic core collapse.
\item There exists a so-far undetected diffuse background of supernova neutrinos, produced
by all past supernovae occurring in the universe.  Future detectors that may approach
the megaton scale should be able to see a few events from this source.  Detection of
these neutrinos would place a important constraint on the inventory of massive stars
undergoing core collapse, from the first epoch of star formation until now.
\item Supernovae are one of the most important engines for nucleosynthesis, controlling
much of the chemical enrichment of the galaxy.  As described in the next section,
neutrinos are directly and indirectly involved in this synthesis.
\item While most of the energy released in core collapse is radiated as neutrinos over the
first several seconds, neutrino emission at a lower level continues as the proto-neutron
star cools and radiates away its lepton number.  It is quite possible that phase changes
in the dense nuclear matter could occur several tens of seconds after core bounce,
altering the late-time neutrino ``light curve" in a characteristic way.   
While there are considerable uncertainties in
estimates, neutrino processes may continue to dominate neutron star cooling for
$\sim 10^5$ years.
\item  The neutrino burst could include other sharp features in time, marking interesting
astrophysics.  The melting of iron to nucleons with the passage of the shock wave through
the outer iron core is predicted to produce a spike in the neutrino luminosity, lasting for
a few milliseconds.  Continued accretion onto the neutron star surface could produce a
collapse to a black hole, and consequently a sudden termination in neutrino emission.
\item  Supernova cooling times place constraints on new physics associated with particles
that also couple weakly to matter.  For example, a light scalar called the axion could,
in principle, compete with neutrinos in cooling a supernovae.  The requirement that axion
emission not shorting the cooling time too much, which would be in conflict with SN1987A
data, constrains the mass and coupling of this hypothesized particle.
\end{itemize}

\section{Neutrinos and Nucleosynthesis}
Neutrinos and nucleosynthesis are both associated with explosive environments 
found in astrophysics.  This section discusses three examples, the Big Bang, the
neutrino process, and the r-process.

\subsection{Neutrinos in Big-Bang Nucleosynthesis (BBN)}
One of the classic problems in nucleosynthesis and cosmology is accounting for
the mass fraction of primordial helium of 25\%, as well as the
abundances of D, ${}^3$He,  and ${}^7$Li.  The ${}^4$He mass fraction
shows that nuclei were synthesized from the
early-universe nucleon soup at a time when the n/p ratio was $\sim$ 1/7.   This requires
the proper coordination of two ``clocks," one being the weak interaction rate for decay of
free neutrons to protons, and the other the Hubble rate governing the expansion of the
universe,
\[ H(t) \equiv {1 \over R(t)} {d R(t) \over dt} = \sqrt{ 8 \pi G \rho(t) \over 3}, \]
where $G$ is the gravitational constant.
This clock depends on the energy density $\rho(t)$ which, at this epoch, is 
dominated by relativistic particles, including the neutrinos.   BBN
also includes one adjustable parameter, the baryon density, which is usually given
in terms of the ratio of baryons to photons $\eta$.  But the primordial abundance of ${}^4$He is
relatively insensitive to $\eta$, in contrast to other BBN species like D.  Furthermore
an independent and quite precise determination of $\eta$ has now been derived
from the pattern of acoustic peaks seen in mappings of the cosmic microwave
background.  Consequently the abundance of ${}^4$He is a rather good test of the
number of neutrino flavors contributing to early-universe expansion.

Detail analyses find that a consistent picture emerges.  The number of neutrino
species is found to be $N_\nu = 2.4 \pm 0.4$.  One can adopt the expected
value of $N_\nu = 3$ and reproduce the abundances of both D and ${}^4$He at
the 68\% confidence level \cite{steigman}.

This standard use of BBN is now mostly of historical interest, though at one time it was the
best test of the number of light neutrino species.  However, there are
variations of the BBN analysis that are of significant current interest, such as the consequences of
a net lepton number asymmetry in the universe, or the presence of a sterile neutrino
(a neutrino lacking standard-model couplings) that might mix with active species.  Such
phenomena affect resulting abundances, and thus can be constrained in a BBN
analysis.

\subsection{The Neutrino Process}
One of the more amusing roles for neutrinos in nucleosynthesis is found in
the neutrino process, the direct synthesis of new elements through neutrino reactions.
Core-collapse supernovae provide the enormous neutrino fluences
necessary for such synthesis to be significant.  They also eject newly 
synthesized material into the interstellar medium, where it can be incorporated
into a new generation of stars.

Among the elements that might be made primarily or partially in the $\nu$-process,
the synthesis of ${}^{19}$F is one of the more interesting examples \cite{woosley}.
The only stable isotope of fluorine,
${}^{19}$F has an abundance
\[ {^{19}\mathrm{F} \over ^{20}\mathrm{Ne}} \sim {1 \over 3100}. \]
Ne is one of the hydrostatic burning products in massive stars, produced in
great abundance and ejected in core-collapse supernovae.  Thus a mechanism
that converts $\sim$ 0.03\% of the ${}^{20}$Ne in the star's mantle into ${}^{19}$F could
account for the entire observed abundance of the latter.

The Ne zone in a supernova progenitor star
is typically located at a radius of $\sim$ 20,000 km.  A simple calculation
that combines the neutrino fluence through the Ne zone
with the cross section for inelastic neutrino scattering off ${}^{20}$Ne shows that
approximately 0.3\% of the ${}^{20}$Ne nuclei would interact with the
neutrinos produced in the core collapse.  Almost all of these
reactions result in the production of ${}^{19}$F, e.g., 
\begin{eqnarray}
 {}^{20}\mathrm{Ne}(\nu,\nu')^{20}\mathrm{Ne}^* &\rightarrow& {}^{19}\mathrm{Ne} + n 
\rightarrow ^{19}\mathrm{F} + e^+ + \nu_e + n \nonumber \\
  {}^{20}\mathrm{Ne}(\nu,\nu')^{20}\mathrm{Ne}^* &\rightarrow& {}^{19}\mathrm{F}
+ p, \nonumber
\end{eqnarray}
with the first reaction occurring half as frequently as the 
second.  Thus one would expect the abundance
ratio to be ${}^{19}\mathrm{F}/ {}^{20}\mathrm{Ne} \sim 1/ 300$, corresponding to an order
of magnitude more ${}^{19}$F than found in nature.  

This example shows that stars are rather complicated factories for nucleosynthesis.
Implicit in the reactions above are mechanisms that also destroy ${}^{19}$F.  
For example, about 70\% of the neutrons coproduced with ${}^{19}$F in the first reaction
immediately recapture on ${}^{19}$F, destroying the product of interest.  
Similarly, many of the coproduced protons destroy ${}^{19}$F via ${}^{19}$F(p,$\alpha)^{16}$O --
unless the star is rich in ${}^{23}$Na, which readily consumes protons via
${}^{23}$Na(p,$\alpha)^{20}$Ne.  Finally, some of the ${}^{19}$F produced in the
neon shell is destroyed when the shock wave passes through that zone:
the shock wave can heat the inner portion of the Ne zone above $1.7 \times 10^9$K, the
temperature at which ${}^{19}$F can be destroyed by ${}^{19}$F($\gamma,\alpha)^{15}$N.

If all of this physics is treated carefully in a nuclear network code, one finds that the
desired ${}^{19}$F/${}^{20}$Ne $\sim$ 1/3100 is achieved for a heavy-flavor neutrino
temperature of about 6 MeV.  This is quite consistent with the temperatures that come
from supernova models.

The neutrino process produces interesting abundances of several relatively
rare, odd-A nuclei including ${}^7$Li, ${}^{11}$B, 
${}^{138}$La, ${}^{180}$Ta, and ${}^{15}$N.  Charged-current neutrino reactions
on free protons can produce neutrons that, through $(n,p)$ and $(n,\gamma)$
reactions, lead to the nucleosynthesis of the so-called ``p-process" nuclei from
A=92 to 126.  The production of such nuclei has been a long-standing puzzle in nuclear astrophysics.

\subsection{The r-process}
Beyond the iron peak nuclear Coulomb barriers become so high
that charged particle reactions become ineffective, leaving
neutron capture as the mechanism responsible for producing
the heaviest nuclei.
If the neutron abundance is modest,
this capture occurs in such a way that each newly synthesized
nucleus has the opportunity to $\beta$ decay, if it is energetically
favorable to do so.  Thus weak equilibrium is maintained within
the nucleus, so that synthesis is along the path of stable 
nuclei.  This is called the s- or slow-process.  However a
plot of the s-process in the (N,Z) plane reveals that this
path misses many stable, neutron-rich nuclei that are known to
exist in nature.  This suggests that another mechanism is at
work, too.  Furthermore, the abundance peaks found in nature 
near masses A $\sim$ 130 and A $\sim$ 190, which mark the closed
neutron shells where neutron capture rates and $\beta$ decay
rates are slower, each split into two subpeaks.  One set of subpeaks
corresponds to the closed-neutron-shell numbers N $\sim$ 82
and N $\sim$ 126, and is clearly associated with the s-process.
The other set is shifted to smaller N $\sim$ 76 and $\sim$ 116,
respectively, suggestive of a much more explosive
neutron capture environment. 
  
This second process is the r- or rapid-process \cite{qian}, which requires a neutron fluence
so large that neutron capture is fast compared to $\beta$ decay.  In this case
nuclei rapidly absorb neutrons until they approach the neutron drip line.  That is,
equilibrium is maintained by $(n,\gamma) \leftrightarrow
(\gamma,n)$, not by weak interactions.  Consequently the nuclei participating in the r-process are very
different from ordinary nuclei -- very neutron rich nuclei that would decay immediately
in the low-temperature environment of Earth.   The rate of nucleosynthesis is
controlled by the rate of $\beta$ decay: a new neutron can be captured only
after $\beta$ decay, $n \rightarrow p + e^- +\bar{\nu}_e$,
opens up a hole in the neutron Fermi sea.  Consequently one expects abundance
peaks near the closed neutron shells at N $\sim$ 82 and 126, as $\beta$ decay 
is slow and mass will pile up at these ``waiting points."  By a series of rapid neutron captures 
and slower $\beta$ decays, synthesis can proceed all the way to the transuranics.  This,
of course, requires the very explosive conditions necessary to support large neutron
densities (typically $\rho(n) \sim 10^{18}-10^{20}$/cm$^3$, temperatures $\sim$
10$^9$ K), sufficient time ($\sim$ 1 second), and a ratio of neutrons to heavy seed 
nuclei of $\gsim$ 100 (so that
there are enough neutrons to add to each seed nucleus).
  
The path of the r-process is along neutron-rich nuclei, 
where the neutron Fermi sea is just $\sim$ (2-3) MeV away from
the neutron drip line (where no more bound neutron levels exist).
After the r-process finishes (the neutron exposure ends)
the nuclei decay back to the valley of stability by $\beta$
decay.  This involves conversion of neutrons into protons,
and that shifts the r-process peaks from the parent-nucleus values of N $\sim$ 82 and 126
to lower values for the stable daughter nuclei, off course.  This effect is clearly seen in the
abundance distribution -- a very strong hint that many familiar heavy
nuclei came from parents created in exotic, explosive environments.

\begin{figure}
\begin{center}
\includegraphics[width=14cm]{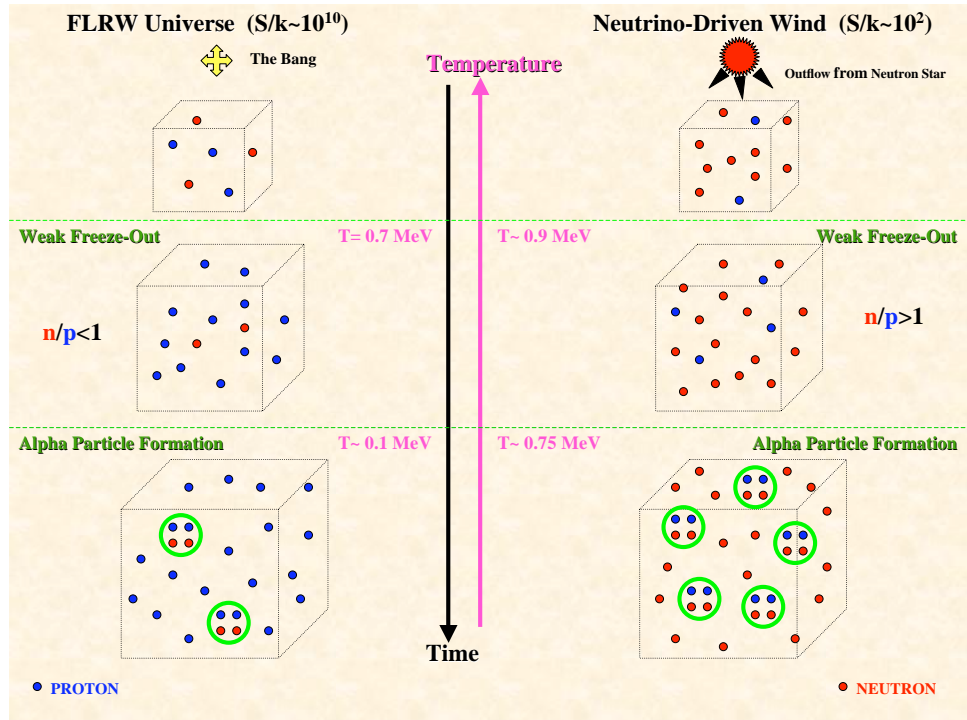}
\end{center}
\caption{ A comparison of two remarkably similar mechanisms for nucleosynthesis,
an expanding, cooling, high entropy, proton-rich nucleon gas (the Big Bang, left panel)
and an expanding, cooling, high entropy, neutron-rich nucleon gas (the supernova wind r-process,
right panel).  In the former, the synthesis
determinates at $^4$He and a few other light elements. In the latter, the
synthesis continues to the heavy transuranic elements.  Figure from G. Fuller.}
\label{fig:rprocess}
\end{figure}

The role of neutrinos in the r-process is in maintaining that
explosive, neutron-rich environment: while there are alternatives,
perhaps the most plausible site is the neutrino-driven wind -- the
last ejecta blown off the neutron star -- of core-collapse supernovae.
This material is a hot,
radiation-dominated gas containing neutrons
and protons, with an excess of neutrons.  As the nucleon gas expands
off the star and cools, it goes through
a freezeout to $\alpha$ particles, a step that essentially
locks up all the protons in this way.
Then the $\alpha$s interact through reactions like 
\[ \alpha + \alpha +\alpha \rightarrow ^{12}C  \]
to start forming heavier nuclei.  The
$\alpha$ capture continues, eventually synthesizing intermediate-mass
``seed" nuclei.  Once these seed nuclei are produced, if the requisite number of
neutrons is available ($\sim$ 100 per seed nucleus) the synthesis of very heavy nuclei
is possible.   The scenario, as depicted in Fig.~\ref{fig:rprocess},
is quite similar to the Big Bang:  a hot nucleon gas (with an entropy
$S\sim$ 100 in Boltzmann units, compared to the BBN $S\sim$ 10$^{10}$)
expand and cools, condensing into nuclei.  But a detail -- the neutrino
wind has an excess of neutrons, while the Big Bang is proton-rich -- leads
to the synthesis of uranium in one case, and to termination of nucleosynthesis
at $^4$He (plus a few light elements) in the other.
Some of the supernova
modelers find conditions where a neutrino-wind r-process almost happens.  The
neutrinos are crucial: they help keep the entropy of the nucleon
gas high, control the n/p ratio
of the gas through competing charge-current reactions, and generate the wind
that ejects the r-process products.

There are some very nice aspects of this site: the amount of
matter ejected is $\sim$ 10$^{-5} - 10^{-6}$ solar masses,
a production per event that if integrated over the lifetime of the
galaxy gives the required total abundance of r-process metals,
assuming typical supernova rates.  There are also
a few problems, especially the fact that with calculated entropies,
the neutron fractions
in the nucleon soup above the proto-neutron star
appear too low to produce a successful A $\sim$ 190 peak.
This may reflect our incomplete understanding of the supernova
mechanism, or possibly our inability to model the weak interactions
(including the effects of neutrino oscillations) adequately.

\section{Neutrino Cooling and Red Giants}
Several neutrino cooling scenarios have already been discussed,
including cooling of the proto-neutron star produce in core collapse 
and cooling connected with the expansion of the early universe.
Red giant cooling provides an additional example of the use of 
astrophysical arguments to constrain
fundamental properties of neutrinos.

\subsection{Red Giants and Helium Ignition}
In a solar-like star, when the hydrogen in the central core has been
exhausted, an interesting evolution ensues:
\begin{itemize}
\item  With no further means of producing energy, the core
slowly contracts, thereby increasing in temperature as gravity
does work on the core.
\item Matter outside the core is still hydrogen rich, and
can generate energy through hydrogen burning.  Thus a hydrogen-burning
shell forms, generating the gas pressure supporting the outside layers of the
star.  As the ${}^4$He-rich core contracts, the matter outside the
core is also pulled deeper into the gravitational potential.
Furthermore, the H-burning shell continually adds more mass to the core.
This means the burning in the shell must intensify to generate
the additional gas pressure to fight gravity.  The shell also
thickens, as more hydrogen is above the
burning temperature.
\item The resulting increasing gas pressure causes the outer
envelope of the star to expand by a large factor, up to a 
factor of 50.  The increase in radius more than compensates for
the increased internal energy generation, so that a cooler
surface results.  The star reddens.  Stars of this type are
called red supergiants.
\item This evolution is relatively rapid, perhaps a few
hundred million years: the dense core requires large energy
production.  The helium core is supported
by its degeneracy pressure, and is characterized by densities
$\sim 10^6$ g/cm$^3$.  This stage ends when the
core reaches densities and temperatures that allow helium burning
through the reaction
\[ \alpha + \alpha + \alpha \rightarrow ^{12}C + \gamma . \]
As this reaction is quite temperature dependent,
the conditions for ignition are very sharply defined.
This has the consequence that the core mass at the helium flash point
is well determined.
\item  The onset of helium burning produces a new source of
support for the core.  The energy released elevates the temperature
and the core expands: He burning, not electron degeneracy, now 
supports the core.  The burning shell and envelope move
outward, higher in the gravitational potential.  Thus shell
hydrogen burning slows (the shell cools) because less gas pressure
is needed to satisfy hydrostatic equilibrium.  All of this
means the evolution of the star has now slowed: the red giant
moves along the ``horizontal branch," as interior temperatures
increase slowly, much as in the main sequence.
\end{itemize}

The 3$\alpha$ process involves some fascinating nuclear physics
that will not be recounted here:  the existence of certain nuclear resonances
was predicted based on the astrophysical requirements for this process.
The resulting He-burning rate exhibits a sharp temperature dependence
$\sim$ T$^{40}$ in the range relevant to red giant cores.
This dependence is the reason the He flash
is delicately dependent on conditions in the core.

\subsection{Neutrino Magnetic Moments and He Ignition}
Prior to the helium flash, the degenerate He core radiates
energy largely by neutrino pair emission.  The process is
the decay of a plasmon --- which one can think of as a photon
``dressed" by electron-hole excitations --- thereby acquiring 
an effective mass of about 10 keV.  The photon couples to
an electron particle-hole pair that
then decays via $Z_o \rightarrow \nu \bar{\nu}$. 

If this cooling is somehow enhanced, the degenerate helium core 
would not ignite at the normal
time, but instead continue to grow.    When the core does finally
ignite, the larger core will alter
the star's subsequent evolution.

One possible mechanism for enhanced cooling is a neutrino
magnetic moment.  Then the plasmon could directly couple to
a neutrino pair.  The strength of this coupling would 
depend on the size of the magnetic moment.

A delay in the time of He ignition has several observable
consequences, including changing the ratio of red giant to
horizontal branch stars.  Thus, using the standard theory of
red giant evolution, investigators have attempted to determine
what size of magnetic moment would produce unacceptable 
changes in the astronomy.  The result is a limit \cite{raffelt} on the
neutrino magnetic moment of
\[ \mu_{ij} \lsim 3 \times 10^{-12} \mathrm{~electron~Bohr~magnetons} .\]
This limit is two orders of magnitude more stringent
than that obtained from direct laboratory tests, e.g., experiments looking 
for the effect of a neutrino magnetic moment in the scattering of
reactor neutrinos off electrons.

This example is just one of a number of such constraints that
can be extracted from similar stellar cooling arguments.
The arguments above, for example, can be repeated for 
neutrino electric dipole moments, or for
axion emission from red giants.   As noted previously, the
arguments can be extended to supernovae: anomalous
cooling processes that shorten the cooling time in a way that is
inconsistent with SN1987A observations are ruled out.  
Limits can also be placed on Dirac neutrino masses:
the mass term would allow neutrinos to scatter into sterile right-handed states,
which would then immediately escape, carrying off energy.

\section{High Energy Astrophysical Neutrinos}
\begin{figure}
\begin{center}
\includegraphics[width=12cm]{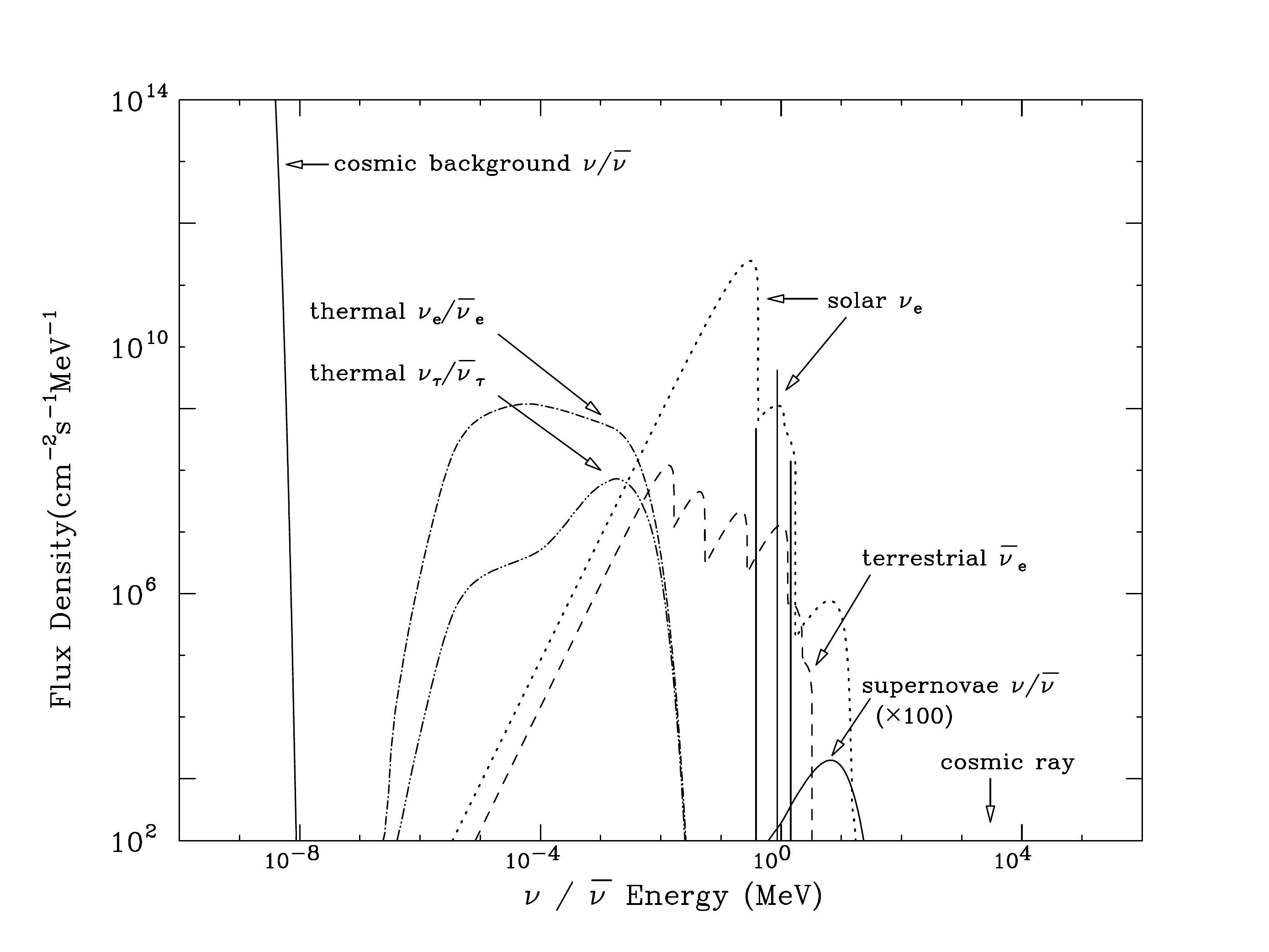}
\end{center}
\caption{The low-energy neutrino ``sky."  In addition to the principal sources
discussed in the text (Big-Bang, solar, supernova, and atmospheric neutrinos), the
natural neutrino background includes $\sim$ 1 keV thermal neutrinos of all flavors emitted
by the sun as well as $\bar{\nu}_e$s generated by the Earth's natural radioactivity.
From \cite{lin}.}
\label{fig:lowEnus}
\end{figure}

\begin{figure}
\begin{center}
\includegraphics[width=14cm]{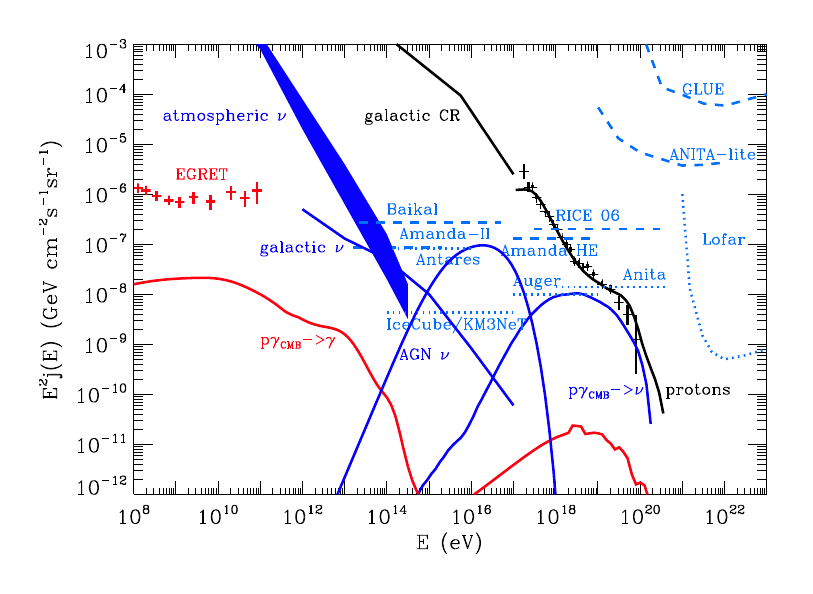}
\end{center}
\caption{A theoretical model of high-energy neutrino ``sky."  Shown are the expected
neutrino fluxes (blue area and lines), the primary cosmic ray fluxes (determined from data
and a model, black), and the secondary $\gamma$-ray fluxes expected from protons interacting
with the microwave background (red).  The neutrino flux is per flavor and includes
only relatively certain sources: atmospheric neutrinos, neutrinos resulting from cosmic-ray
interactions in our galaxy, and GZK neutrinos resulting from cosmic-ray interactions
with the microwave and infrared background.  The figure includes experimental data,
limits, and projected sensitivities to existing and planned telescopes.  Figure from G. Sigl \cite{sigl}.}
\label{fig:highEnusnew}
\end{figure}

The previous discussion focused on astrophysical neutrino sources
with energies ranging from the cosmic microwave/neutrino temperature to $\lsim$ 10 GeV, 
which includes the bulk of atmospheric neutrinos.  These sources are displayed in Fig.~\ref{fig:lowEnus}
according to their contributions to the terrestrial flux density.  This figure includes sources,
such as the thermal solar neutrinos of all flavors, not explicitly discussed here
because of space limitations.  However, the spectrum
of neutrinos is believed to extend to far higher energies due to neutrino production in
some of nature's most spectacular natural accelerators.  The program of
experiments to map out the high-energy neutrino spectrum and its astrophysical sources
is just beginning.   Some guidance is provided by existing data on cosmic-ray protons, nuclei, and $\gamma$ rays, which constrain possible neutrino fluxes (see Fig.~\ref{fig:highEnusnew}). This concluding chapter discusses some of the suggested sources and current efforts 
to develop high-energy neutrino telescopes appropriate to these
sources.  The high-energy spectrum is one of the frontiers of neutrino astronomy.

\subsection{Neutrino Production by Cosmic Rays}
The ultra-high-energy cosmic ray (UHECR) spectrum -- presumably protons and nuclei -- is known to
vary smoothly up to an energy $E$ $\sim$ 4 $\times$ 10$^{19}$ eV.  The spectrum just below
this point is characterized by a spectral index $\alpha \sim -2.7$: the flux varies as
$E^{\alpha}$ \cite{pierre}.  Higher energy events
are seen, but the flux  drops off steeply beyond this point.  This is consistent with the
prediction of Greisen, Zatsepin, and Kuz'min \cite{GZK} of a cutoff in the spectrum above $\sim$ 4 $\times$ 10$^{19}$ eV.  Above this cutoff UHECRs can loose energy by scattering off
microwave photons, producing pions.  This sharply reduces 
the mean-free path of such UHECRs.  The flux drops, reflecting the reduced number of sources
within a mean-free path of the Earth.

An estimate can be made of the flux of UHE neutrinos
associated with the decays of pions and other secondaries produced in the GZK
scattering.  Uncertainties in this estimate include the flux, spectrum, and composition
of the UHECRs, the behavior of the spectrum beyond the GZK cutoff (as
we are blind to these UHECRs), and the spectrum's cosmological evolution.  One bound was obtained by Bahcall and Waxman \cite{waxman}; another is shown
in Fig.~\ref{fig:highEnusnew}.

These uncertainties are connected to some very interesting astrophysical issues:
the maximum energies that can be reached in
astrophysical accelerators; the UHECR uniformity over time (or equivalently redshift); and the role
other background photon sources, such as the infrared and optical backgrounds, in
producing UHE neutrinos.  A
more extended assumption can be found in \cite{APS}.

As in the case of low-energy sources such as solar neutrinos, the detection of very high
energy astrophysical neutrinos would open up new opportunities in both astrophysics
and particle physics.  Because the GZK cutoff limits the range of the UHECRs, neutrinos
provide the only direct probe of nature's most energetic astrophysical accelerators.  
Because neutrinos travel in straight lines through magnetic fields, they point back to
their sources, allowing astronomers to correlate those sources with their optical
counterparts -- the accretion disks surrounding supermassive black holes, quasars, $\gamma$-ray bursts, etc.  The interactions of such energetic neutrinos with matter are untested experimentally,
as terrestrial accelerators have reach only the TeV scale.

\subsection{Point Sources and Neutrino Telescopes}
The possibility of point sources is generally considered
 the astrophysical ``driver" for developing instruments to measure the highest
energy neutrinos.   There are intensely energetic sources in the sky, including 
active galactic nuclei (AGNs),
supernovae and phenomena like $\gamma$-ray bursts that may be associated with
particularly energetic supernovae, and compact objects such as black holes and neutron
stars.    The magnetic fields, shock waves, gravitational fields, and energy densities associated
with such objects are
beyond those that can be produced in the laboratory.  

\begin{figure}
\begin{center}
\includegraphics[width=12cm]{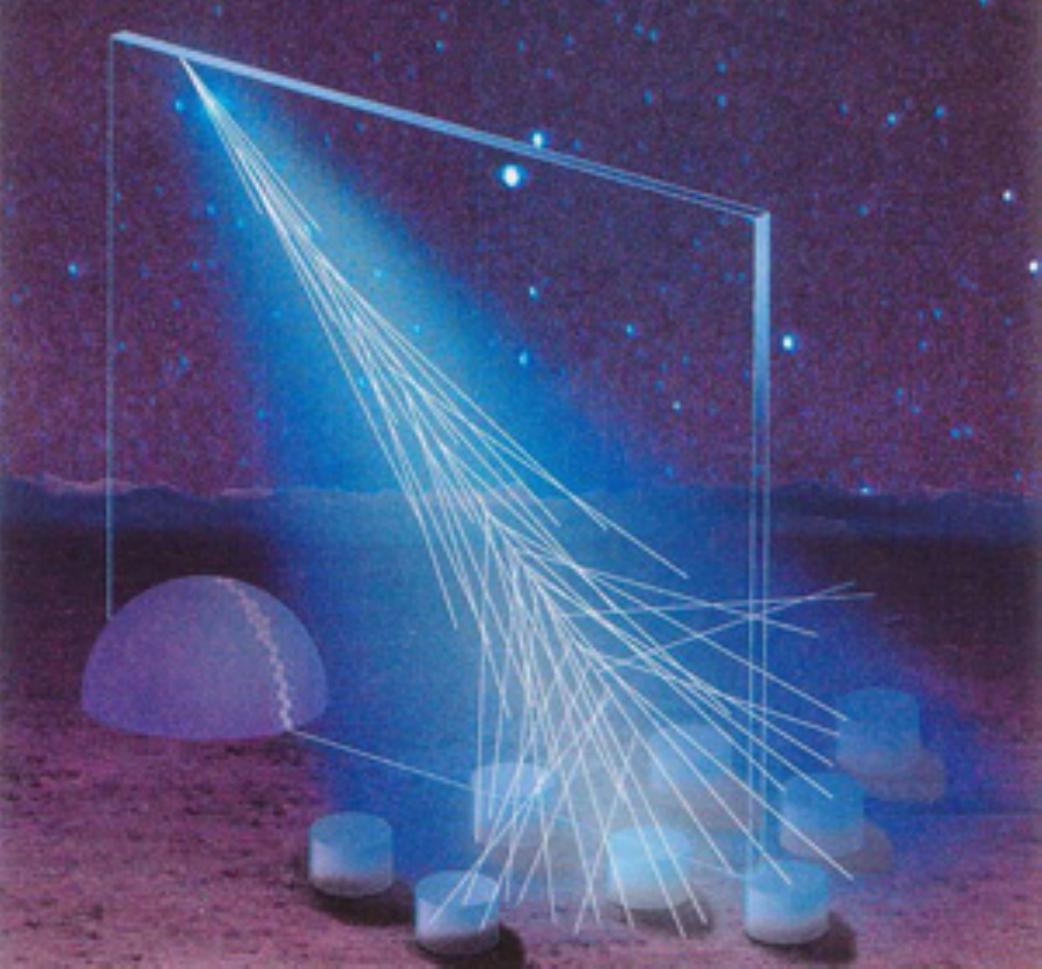}
\end{center}
\caption{A conceptual drawing of the Pierre Auger cosmic ray observatory, which
combines an array of surface detectors (tanks containing water)  with air-florescence telescopes
(which track the development of air showers through the ultraviolet light they produce) to better 
determine the energy of UHECRs.  Figure coutesy of the Pierre Auger Observatory.}
\label{fig:pierre}
\end{figure}

There are recent results that further motivate such neutrino studies.  The cosmic-ray telescope
Pierre Auger (see Fig.~\ref{fig:pierre}), in its studies of events near the GZK cutoff, has found
correlations between clusters of events and nearby AGNs:  At these energies the trajectories
of protons and nuclei are not strongly perturbed by magnetic fields, so that these also
point back to their sources \cite{pierre}.   Once one can attribute events to an astrophysical
source, then those events become a probe of that object.  In this particular case, one is
then challenged to explain the mechanism by which an AGN accelerates nucleons or nuclei to,
and perhaps beyond, $\sim$ 10$^{20}$ eV. 
Were neutrinos also seen, the relative yields of high-energy nuclei and neutrinos would
place important constraints on the astrophysical accelerator.
Similarly, $\gamma$-ray bursts, the most luminous electromagnetic events found in the
universe, appear to be highly beamed emissions of a few seconds in duration.  A few
such events are detected each week.  One
popular model associates these emissions with jets produced in the collapse of the core
of a very massive star into a black hole.   Detection (or failure to detect) neutrinos from
such a source could help modelers determine whether the plasmas accelerated by such jets
contain a significant density of nucleons and nuclei, as photonuclear excitation 
would necessarily lead to
neutrinos.

\begin{figure}
\begin{center}
\includegraphics[width=12cm]{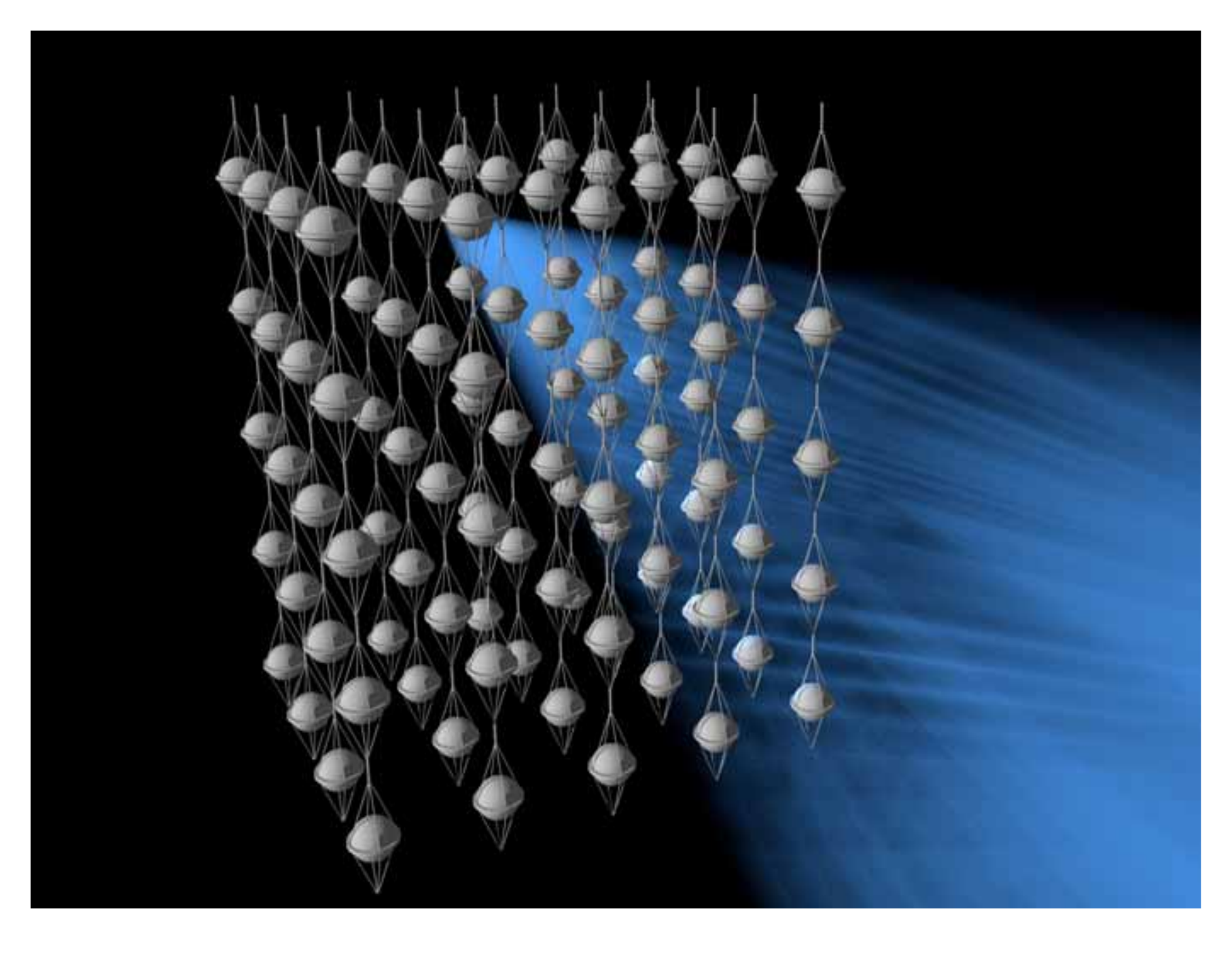}
\end{center}
\caption{Artist's rendering of the IceCube detector, showing the optical modules arrayed
on strings within the deep ice and a cone of blue Cerenkov light produced by a
neutrino scattering within the detector.  Figure courtesy of the IceCube Project, 
University of Wisconsin, Madison \cite{halzen}.}
\label{fig:icecube}
\end{figure}

The challenge in the field of UHE neutrinos is to build telescopes with the necessary
sensitivity to see events, given current estimates of the fluxes (see Fig.~\ref{fig:highEnusnew}).
This requires instrumenting very large volumes.   There have been ongoing efforts
to use large quantities of water and ice as detectors, with experiments developed or planned in
lakes (Baikal), in the Antarctic (AMANDA), and in oceans (NESTOR, ANTARES) \cite{APS}.
IceCube \cite{halzen}, a project that will extend the dimensions of such detectors to a cubic
kilometer, is now under construction at the South Pole (Fig.~\ref{fig:icecube}).   This
telescope, when completed, will view the ice through approximately 4200  optical modules,
deployed on 70 vertical strings at a depth of 1450 to 2450m.  At this depth the clarity of
the ice enhances event detection.  This deep detector is coupled to a surface air-shower
array.  Approximately half of the strings have been deployed; the completion of the
entire array is expected by 2011.

\section{Acknowledgement}
This work was supported in part by the U.S. Department of Energy, Office of Nuclear Physics,
under grant \#DE-FG02-00ER41132.


\begin{thebibliography}{000}

\bibitem{hh00} W. C. Haxton and B. R. Holstein, Am. J. Phys. {\bf 68} (2000) 15 and
{\bf 72} (2004) 18.

\bibitem{bahcallbook} J. N. Bahcall, {\em Neutrino Astrophysics}, (Cambridge University,
Cambridge, 1989).

\bibitem{davis} R. Davis Jr., D. S. Harmer, and K. C. Hoffman, Phys. Rev. Lett. {\bf 20} (1966) 1205.

\bibitem{BSB} J. N. Bahcall, A. Serenelli, and S. Basu, Ap. J. {\bf 621} (2005) L85.

\bibitem{TC} A. S. Brun, S. Turck-Chieze, and P. Morel, Ap. J. {\bf 506} (1998) 913.

\bibitem{BP} J. N./ Bahcall and M. H. Pinsonneault, Phys. Rev. Lett. {\bf 92} (2004) 121301.

\bibitem{SKII}  J. Hosaka {\it et al.}, Phys. Rev. {\bf D73} (2006) 112001; J. P. Cravens {\it et al.},
arXiv:0803.4312 (2008).

\bibitem{SNOI} Q. R. Ahmad {\it et al.}, Phys. Rev. Lett. {\bf 89} (2002) 011301.

\bibitem{SNOsn} Q. R. Ahmad {\it et al.}, Phys. Rev. {\bf C75} (2007) 045502.

\bibitem{MS} S. P. Mikheyev and A. Smirnov, Sov. J. Nucl. Phys. {\bf 42} (1985) 913.

\bibitem{Wolf} L. Wolfenstein, Phys. Rev. {\bf D17} (1979) 2369.

\bibitem{SNOKam} B. Aharmim {\it et al.}, arXiv:0806.0989 (submitted to Phys. Rev. Lett.);
KamLAND Collaboration, Phys. Rev. Lett. {\bf 92} (2005) 081801.

\bibitem{Borexino} Borexino Collaboration, Phys. Lett. {\bf B658} (2008) 101.

\bibitem{IMB} T. J. Haines {\it et al.}, Phys. Rev. Lett. {\bf 20} (1986) 1986;
D. Casper {\it et al.}, Phys. Rev. Lett. {\bf 66} (1991) 2561.

\bibitem{Kamioka}  K. S. Hirata {\it et al.}, Phys. Lett. {\bf B205} (1988) 416.

\bibitem{SKatmos} Y. Fukuda {\it et al.}, Phys. Rev. Lett. {\it 81} (1998) 1562. 

\bibitem{APS} {\em APS Mult-Divisional Study of the Physics of Neutrinos},
http://www.aps.org/policy/reports/multidivisional/neutrino/. 

\bibitem{models} H.-Th. Janka, K. Langanke, A. Marek, G. Martinez-Pinedo, and B. Muller, Phys. Rep. {\bf 442} (2007) 38; A. Mezzacappa, Ann. Rev. Nucl. Part. Sci. {\bf 55} (2005) 467;
K. Kotake, K. Sato, and K. Takahashi, Rep. Prog. Phys. {\bf 69} (2006) 971;
S. Woosley and J. S. Bloom, Ann. Rev. Astron. Astrophys. {\bf 44} (2006) 507.

\bibitem{snnus} K. Hirata {\it et al.}, Phys. Rev. Lett. {\bf 58} (1987) 1490; R. M. Bionta {\it et al.},
Phys. Rev. Lett. {\bf 58} (1987) 1494.

\bibitem{fuller} G. M. Fuller, R. W. Mayle, J. R. Wilson, and D. N. Schramm, Ap. J. {\bf 322} (1987) 795;
D. Notzold and G. Raffelt, Nucl. Phys. {\bf B307} (1988) 924.

\bibitem{steigman} G. Steigman, Ann. Rev. Nucl. and Part. Sci. {\bf 57} (2007) 463.

\bibitem{woosley} S. E. Woosley, D. H. Hartmann, R. D. Hoffman, and W. C. Haxton, 
Ap. J. {\bf 356} (1990) 272.

\bibitem{qian} Y.-Z. Qian, Prog. Part. Nucl. Phys. {\bf 50} (2003) 153.

\bibitem{raffelt} G. G. Raffelt, Phys. Rev. Lett. {\bf 64} (1990) 2856.

\bibitem{lin} W.C. Haxton and W. Lin, Phys. Lett. {\bf B486} (2000) 263.

\bibitem{sigl} G. Sigl, Nuc. Phys. Proc. Suppl. {\bf 168} (2007) 219.

\bibitem{pierre} The Pierre Auger Collaboration, arXiv:0806.4302 (to be published in Phys.
Rev. Lett.)

\bibitem{GZK} K. Greisen, Phys. Rev. Lett. {\bf 16} (1966) 748; G. T. Zatsepin and V. A.
Kuz'min, JETP Lett. {\bf 4} (1966) 78.

\bibitem{waxman} E. Waxman and J. N. Bahcall, Phys. Rev. {\bf D59} (1999) 023002.

\bibitem{halzen} J. Ahrens {\it et al.}, Astropart. Phys. {\bf 20} (2004) 507;
F. Halzen, Eur. Phys. J. {\bf C46} (2006) 669; E. Resconi {\it et al.}, arXiv:0807:3891.


\end{thebibliography}
\end{document}